\RequirePackage{rotating}
\documentclass[%
aip,
jcp,
amsmath,amssymb,
preprint,%
]{revtex4-1}

\setcitestyle{numbers,square,sort&compress}

\usepackage{graphicx}
\usepackage{dcolumn}
\usepackage{bm}
\usepackage{CJKutf8}
\graphicspath{{figures/}}
\usepackage[utf8]{inputenc}
\usepackage[T1]{fontenc}
\usepackage{mathptmx}
\usepackage{siunitx}
\usepackage{tabularx}
\usepackage{threeparttable,booktabs}
\usepackage[version=3]{mhchem} 
\usepackage{xcolor}
\usepackage{tikz}

\newcommand{\invis}[1]{{\color{white}#1}}
\newcommand{\ee}[1]{\ensuremath{\times 10^{#1}}}
\usepackage{cancel}
\usepackage{verbatim}
\usepackage{tablefootnote}
\usepackage{rotating}

\usepackage{changes}
\usepackage{ulem}

\makeatletter
\def\@email#1#2{%
 \endgroup
 \patchcmd{\titleblock@produce}
  {\frontmatter@RRAPformat}
  {\frontmatter@RRAPformat{\produce@RRAP{*#1\href{mailto:#2}{#2}}}\frontmatter@RRAPformat}
  {}{}
}%
\makeatother

\begin{document}
\preprint{AIP/123-QED}
\begin{CJK*}{UTF8}{gbsn}

\title[\ce{C2} predissociation calculation]{Multireference configuration interaction study of the predissociation of \ce{C2} via its $F\,^1\Pi_u$ state}

\author{Zhongxing Xu (徐重行)}
\affiliation{Department of Chemistry, University of California – Davis}

\author{S. R. Federman}
\affiliation{Department of Physics and Astronomy, University of Toledo}

\author{William M. Jackson}
\affiliation{Department of Chemistry, University of California – Davis}

\author{Cheuk-Yiu Ng}
\affiliation{Department of Chemistry, University of California – Davis}

\author{Lee-Ping Wang}
\affiliation{Department of Chemistry, University of California – Davis}

\author{Kyle N. Crabtree}
\affiliation{Department of Chemistry, University of California – Davis}
\email{kncrabtree@ucdavis.edu}

\date{\today}

\begin{abstract}

Photodissociation is one of the main destruction pathways for dicarbon (\ce{C2}) in astronomical environments such as diffuse interstellar clouds, yet the accuracy of modern astrochemical models is limited by a lack of accurate photodissociation cross sections in the vacuum ultraviolet range.
\ce{C2} features a strong predissociative $F\,^1\Pi_u - X\,^1\Sigma_g^+$ electronic transition near 130\,nm originally measured in 1969; however, no experimental studies of this transition have been carried out since, and theoretical studies of the $F\,^1\Pi_u$ state are limited.
In this work, potential energy curves of excited electronic states of \ce{C2} are calculated with the aim of describing the predissociative nature of the $F\,^1\Pi_u$ state and providing new ab initio photodissociation cross sections for astrochemical applications.
Accurate electronic calculations of 56 singlet, triplet, and quintet states are carried out at the DW-SA-CASSCF/MRCI+Q level of theory with a CAS(8,12) active space and the aug-cc-pV5Z basis set augmented with additional diffuse functions.
Photodissociation cross sections arising from the vibronic ground state to the $F\,^1\Pi_u$ state are calculated by a coupled-channel model.
The total integrated cross section through the $F\,^1\Pi_u$ $v=0$ and $v=1$ bands is 1.198\ee{-13}\,cm$^2$cm$^{-1}$, giving rise to a photodissociation rate of 5.02\ee{-10}\,s$^{-1}$ under the standard interstellar radiation field, much larger than the rate in the Leiden photodissociation database.
In addition, we report a new $2\,^1\Sigma_u^+$ state that should be detectable via a strong $2\,^1\Sigma_u^+-X\,^1\Sigma_g^+$ band around 116\,nm.

\end{abstract}

\maketitle

\end{CJK*}

\section{\label{sec:C2_theo_intro}Introduction}

\begin{figure}
 \centering
 \begin{tikzpicture}
  \node[anchor=south west] at (0,0){\includegraphics[width=15cm]{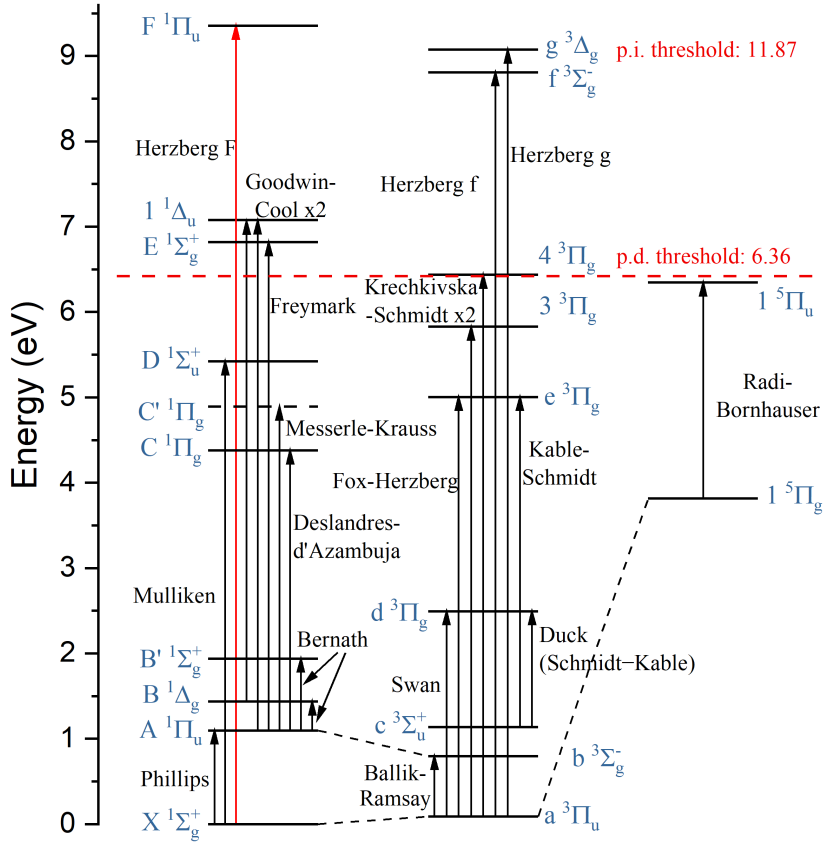}};
 \end{tikzpicture}
 \caption{Electronic states and band systems of \ce{C2}.}\label{fig:states}
\end{figure}

\ce{C2} is an important small molecule that is widely found in hydrocarbon combustion,~\cite{Wollaston1802} comets,~\cite{Donati1864} and astronomical environments.~\cite{Lambert1974,Sonnentrucker2007,Wehres2010}
As a homonuclear diatomic molecule with no dipole-allowed rotational or vibrational transitions, observation of \ce{C2} is accomplished via transitions among its electronic states.
To date, 20 electronic states have been studied by spectroscopy, with a number of low-lying excited states being newly found in recent years.~\cite{McKemmish2020,Kokkin2006,Krechkivska2015,Krechkivska2017,Welsh2017}
In the past 15 years, Schmidt and his colleagues have explored several new electronic states and vibrational levels of \ce{C2} both experimentally and theoretically.~\cite{Schmidt2021}
An overview of the electronic states of \ce{C2} and its observed spectroscopic bands is shown in Figure~\ref{fig:states}.

Owing to its fundamental nature, \ce{C2} has been the subject of a number of theoretical investigations.
It is well known that the $X\,^1\Sigma_g^+$ ground state of \ce{C2} has a multi-reference nature due to the quasi-degeneracy of the $2\,\sigma_u^*$, $1\,\pi_u$, and $3\,\sigma_g$ frontier molecular orbitals (MOs).
This has even led to debate about whether the chemical bond in \ce{C2} is better described as a double-$\pi$ bond, as conventional MO theory predicts, or a quadruple bond on the basis of valence bond theory arguments.~\cite{Shaik2012,Shaik2017}
The near-degeneracy of the frontier MOs is also responsible for the presence of an additional 7 low-lying excited electronic states with energies below 3~eV.
In particular the first triplet $a\,^3\Pi_u$ state lies only 0.089~eV above the ground state, and allowed transitions among the low-lying triplet states (e.g., the Ballik-Ramsay $b\,^3\Sigma_g^- - a\,^3\Pi_u$ band, the Swan $d\,^3\Pi_g - a\,^3\Pi_u$ band, and the Duck (Schmidt-Kable) $d\,^3\Pi_g - c\,^3\Sigma_u^+$ band),  are readily observable in the visible and near-infrared through absorption or fluorescence.~\cite{Amiot1979,Phillips1948,Joester2007}
The singlet manifold contains 3 low-lying excited states, and though the only allowed transition from the ground state is the near-infrared Phillips $A\,^1\Pi_u - X\,^1\Sigma_g^+$ band,~\cite{Ballik1963} the $B\,^1\Delta_g$ and $B^\prime\,^1\Sigma_g^+$ states have been observed via the Bernath bands arising from the $A\,^1\Pi_u$ state.~\cite{Douay1988}
Given the considerable quantity of experimental data available for comparison, these states have been well-characterized by theoretical calculations.~\cite{Shi2011b,Schmidt2007,Kokkin2007}
A recent example is a detailed MRCI+Q/aug-cc-pCV5Z study of the the formation rate of \ce{C2} in collisions of two carbon atoms, which involves all of these low-lying \ce{C2} states.~\cite{Babb2019}

There are two singlet states, three triplet states, and two quintet states below or around the photodissociation limit, which is 6.36~eV.~\cite{Borsovszky2021}
The Mulliken $D\,^1\Sigma_u^+ - X\,^1\Sigma_g^+$ band has been well studied.
Previous experiments\cite{Landsverk1939,Blunt1995,Sorkhabi1997a} and theoretical calculations\cite{Schmidt2007} found that the $D-X$ band favors the $\Delta v=0$ sequence.
The most recent study of the Mulliken band explored its $\Delta v=+2$ sequence involving higher vibrational levels of the $D$ state up to $v=1$1\cite{Krechkivska2018}.
The $D-X$ $(0-0)$ band has been widely observed in space\cite{Hupe2012}.
The Deslandres–d’Azambuja $C\,^1\Pi_g - A\,^1\Pi_u$ and Messerle-Krauss $C^{\prime}\,^1\Pi_g - A\,^1\Pi_u$ bands were only observed in very early studies\cite{Dieke1930,Herzberg1940,Phillips1950,Messerle1967}.
However, recent calculations suggested that the Messerle-Krauss band is actually a part of the Deslandres–d’Azambuja band~\cite{Schmidt2021,Jiang2022}, which is also verified by our calculation here.
Among the three $^3\Pi_u$ states in this region, the $e\,^3\Pi_g$ state was discovered early\cite{Fox1937,Phillips1949}.
The $3\,^3\Pi_g$ and $4\,^3\Pi_g$ states were found recently with the aid of theoretical calculations\cite{Krechkivska2015,Krechkivska2017}.
Experiments involving the quintet states are more challenging because the transition to quintet states from singlet or triplet states are forbidden.
Bornhauser \textit{et al.} used perturbation-facilitated optical-optical double resonance spectroscopy to observe the first transition ($1\,^5\Pi_u - 1\,^5\Pi_g$) between quintet states\cite{Bornhauser2015}.
Another two singlet states $E\,^1\Sigma_g^+$ and $1\,^1\Delta_u$ lie in the UV region.
The $E\,^1\Sigma_g^+$ state was detected through the $E\,^1\Sigma_g^+ - A\,^1\Pi_u$ band\cite{Freymark1950}, and the $1\,^1\Delta_u$ state through in the $1\,^1\Delta_u - B\,^1\Delta_g$ and the two photon $1\,^1\Delta_u - A\,^1\Pi_u$ bands using resonance-enhanced multiphoton ionization (REMPI) spectroscopy\cite{Goodwin1988,Goodwin1989}.

In interstellar space, \ce{C2} was first detected in absorption through the (1-0) band of the Phillips $A-X$ system in the diffuse interstellar medium (ISM) toward Cyg.\ OB2 No.\ 12\cite{Souza1977}, and has since been observed in a wide variety of diffuse cloud sources\cite{Snow1978,Lambert1995,Sonnentrucker2007}.
Because rotational emission is forbidden, the rotational levels of \ce{C2} are metastable and their relative populations are used as a tracer for the local gas kinetic temperature\cite{Snow2006}.
At the low temperatures of diffuse clouds \ce{C2} is unreactive with both H and \ce{H2}, and photodissociation is suggested to be its key destruction pathway\cite{Federman1989}.
Among the states above the dissociation limit shown in Fig.~\ref{fig:states} only the $F\,^1\Pi_u$ state is accessible from the ground $X\,^1\Sigma_g^+$ state, and because all \ce{C2} in diffuse clouds is expected to be in the ground $X\,^1\Sigma_g^+$ state, the $F\,^1\Pi_u$ state is especially important for understanding the chemistry of \ce{C2} in astronomical environments.
In 2017, Welsh \textit{et al.} found that the $v=12$ level of the $e\,^3\Pi_g$ state has a reduced lifetime due to predissociation\cite{Welsh2017}.
Later, the photodissociation of \ce{C2} through the $e\,^3\Pi_g$ state at high vibrational levels was directly observed in a velocity-map imaging experiment\cite{Borsovszky2021}.
This process is predicted to be important for cometary \ce{C2} photodissociation.
However, the transition from the ground $X\,^1\Sigma_g^+$ state to the $e\,^3\Pi_g$ state is forbidden, thus the predissociation through $e\,^3\Pi_g$ is not likely to be a significant route to \ce{C2} photodissociation in the ISM.

To date, the only laboratory spectroscopy of the $F-X$ transition was by Herzberg, Lagerqvist, and Malmberg\cite{Herzberg1969a}, where the $f\,^3\Sigma_g^- - a\,^3\Pi_u$, and $g\,^3\Delta_g - a\,^3\Pi_u$ bands were also detected in the 130--145~nm wavelength region.
The derived spectroscopic constants suggested that the three upper states could be described as Rydberg states due to their similarity with those of low-lying electronic states of \ce{C2+}.
The $F-X$ transition has also been detected in ultraviolet spectra of several diffuse cloud lines of sight, yet a number of discrepancies in line positions and transition intensities remain unresolved\cite{Lambert1995,Kaczmarczyk2000,Sonnentrucker2007,Hupe2012}.
Linewidths of transitions involving individual levels of the $F-X$ system are found to be broadened, confirming that the $F$ state has a lifetime of only $\sim$6~ps likely due to predissociation\cite{Hupe2012}.

From a theoretical standpoint, the most comprehensive treatment of \ce{C2} states in the 7--10~eV region is a 2001 multireference configuration interaction (MRCI) study by Bruna and Grein\cite{Bruna2001}.
Focusing specifically on the $F$ state, they found that it is well-described as a $3s$ Rydberg state that correlates to the $1\,^2\Pi_u$ state of \ce{C2+}; i.e., the configuration is $\sigma_u^2\pi_u^33s$ or [$^2\Pi_u,3s$].
The (0,0) band within the Herzberg $F-X$ system was calculated to have an oscillator strength $f_{00} = 0.098$, which was in good agreement with a value of $f_{00} = 0.10\pm0.01$ inferred from astronomical observations\cite{Lambert1995}.
This oscillator strength is calculated to be the largest among all known electronic transitions of \ce{C2}, being larger by nearly a factor of 2 compared with the $D-X$ Mulliken system.
The adiabatic potential energy curve (PEC) of the $F\,^1\Pi_u$ state features non-adiabatic interactions with the $3\,^1\Pi_u$ and $4\,^1\Pi_u$ states which cause it to support only three bound vibrational levels, two of which have been observed in astronomical spectra\cite{Hupe2012}.
Bruna and Grein estimated a radiative lifetime of $\sim$3~ns for the $F\,^1\Pi_u$ state, which is much longer than the inferred lifetime from the aforementioned linewidth measurements\cite{Bruna2001}.
However, they did not explore potential predissociation mechanisms.
Also, their calculation only included excited states which can be reached via absorption from either the $X\,^1\Sigma_g^+$ or the $a\,^3\Pi_u$ states.
The $^3\Sigma_u^+$ and $^3\Sigma_u^-$ states were not investigated, although those states can be involved in predissociation of the $F$ state through spin-orbit couplings.

Estimates of the photodissociation cross section of \ce{C2} in the Leiden photodissociation and photoionization database\cite{Heays2017} are based on MRCI calculations from the mid-1980s\cite{Pouilly1983} .
Despite great efforts, their computations were severely limited by available computational power at the time.
The calculated electronic energy $T_e$ of the $F\,^1\Pi_u$ state was too large by $\sim$0.8~eV and the calculated oscillator strength for the origin band was only $f_{00} = 0.02$, in considerable disagreement with more recent estimates.
These discrepancies, together with the many new astronomical observations since the theoretical research carried by Bruna and Grein\cite{Bruna2001}, call for a new detailed high-level quantum chemical study, which may improve the estimated photodissociation cross sections and also predict new bands and states that may be targets for further experiments.

Recently, we have investigated the photodissociation of \ce{CS} through high-energy Rydberg states using \textit{ab initio} MRCI methods with a reference space generated by the complete active space self-consistent field (CASSCF) technique\cite{Xu2019}.
By including several Rydberg molecular orbitals into the active space of the CASSCF reference and adding extra diffuse functions to the basis set, the adiabatic PECs of several Rydberg states were obtained successfully.
Photodissociation cross sections were then derived by constructing a coupled system of diabatized states, including non-adiabatic and spin-orbit couplings, and solving the coupled-channel radial Schr\"{o}dinger equation.

Here, we employ similar methods to study the photodissociation of \ce{C2} with a particular focus on the $F\,^1\Pi_u$ state and the $F-X$ electronic transition.
To this end, we have computed the PECs of 57 electronic states, and we explore their potential interactions involving the $F$ state.
Compared to the only two previous theoretical studies of the $F\,^1\Pi_u$ state~\cite{Pouilly1983,Bruna2001}, we use a higher-level computational method and a larger basis set.
Moreover, we carry out a detailed investigation of the predissociative nature of the $F\,^1\Pi_u$ state.
The paper is organized as follows.
Details of our theoretical methods are introduced in Section~\ref{sec:C2_theo_theo}.
The results from \textit{ab initio} calculations are presented in Section~\ref{sec:C2_theo_results}, followed by a discussion of the coupled-channel model and computed photodissociation cross sections and rates in Section~\ref{sec:C2_theo_diss}.
Finally, a summary of work and future perspectives are given in Section~\ref{sec:C2_theo_conclusion}.

\section{Theory and Calculations\label{sec:C2_theo_theo}}

\subsection{Ab initio calculation}\label{subsec:C2_theo_theory_abinitio}

Electronic structure calculations were performed using the MOLPRO 2019.1 quantum chemistry software package\cite{Werner2012,Werner2015}.
Initial electronic states were calculated by the dynamically weighted state-averaged complete active space self-consistent field (DW-SA-CASSCF) method, yielding optimized MOs and configuration state functions\cite{Werner1985,Knowles1985}.
Dynamic electron correlation was then treated by use of internally contracted multireference configuration interaction with single and double excitations from a subset of the DW-SA-CASSCF optimized configurations, and 
the Davidson correction was included in the energy calculations (MRCI+Q)\cite{Werner1988,Knowles1988,Knowles1992}.
PECs were generated from a total of 268 single point calculations spanning internuclear distances from 0.8~\AA\ to 15.0~\AA.
In the bonding region, the points were typically spaced by 0.005~\AA, but in some areas near avoided crossings a smaller grid spacing of 0.001\,\AA\ or 0.002\,\AA\ was employed. 

The full point group of \ce{C2} is $D_{\infty h}$, which cannot be calculated directly in MOLPRO.
Calculations are done at $D_{2h}$ symmetry, which is the largest Abelian subgroup of $D_{\infty h}$.
The irreducible representations of $D_{\infty h}$ up to $\Delta$ map onto those of $D_{2h}$ as follows:
\begin{equation}
 \begin{array}{lll}
\Sigma_g^+ \to A_{g}, & \quad  & \Sigma_u^+ \to B_{1u}  \\
\Sigma_g^- \to B_{1g}, & \quad & \Sigma_u^- \to A_{u} \\
\Pi_g \to (B_{2g}, B_{3g}), & \quad & \Pi_u \to (B_{3u}, B_{2u}) \\
\Delta_g \to (A_{g}, B_{1g}), & \quad & \Delta_u \rightarrow (B_{1u}, A_{u})
 \end{array}
\end{equation}
Here, when referring to the number of orbitals or states of each symmetry in $D_{2h}$, we will list them in the order ($a_g$, $b_{3u}$, $b_{2u}$, $b_{1g}$, $b_{1u}$, $b_{2g}$, $b_{3g}$, $a_u$) consistent with the MOLPRO convention.

For these calculations, Dunning's augmented correlation consistent polarized valence quintuple-zeta Gaussian basis set (aug-cc-pV5Z or aV5Z)\cite{Dunning1989,Kendall1992} was used with the addition of 2 additional $s$-type and 2 additional $p$-type diffuse atomic orbitals per atom.
The final basis set we used can therefore be designated as aug-cc-pV5Z-2s2p or aV5Z-2s2p.
The extra orbitals had exponents of 0.01576 and 0.006304 for $s$-type orbitals and 0.01088 and 0.004352 for $p$-type ones, and were generated by the even tempered method with ratio of a 2.5 from the smallest exponents of the existing basis.
As discussed later, the additional diffuse functions are important for obtaining accurate electronic energies for Rydberg states.
In total, the basis set comprises 270 orbitals, with (50,33,33,19,50,33,33,19) symmetry-adapted functions in $D_{2h}$.
Tests were also performed using additional Dunning's augmented core-valence basis sets aug-cc-pCV5Z and aug-cc-pCV6Z\cite{Dunning1989,Kendall1992}; as expected, for Rydberg states these basis sets had minimal effect on the calculated energy but increased the calculation time by a factor of about 2.5.

\begin{figure}
 \centering
 \begin{tikzpicture}
  \node[anchor=south west] at (0,0) {\includegraphics[width=6cm]{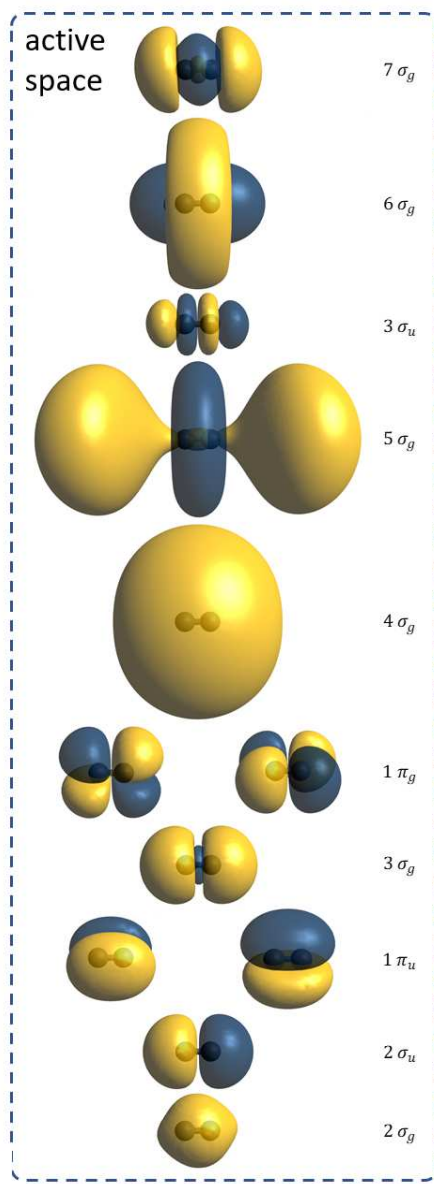}};
 \end{tikzpicture}
 \caption{Molecular orbitals (MOs) in the active space of \ce{C2} optimized at $R=$1.25\,\AA, plotted with isosurface value 0.01. The MOs are generated by a DW-SA-CASSCF calculation with details described in the text except that h-type orbitals are removed from the basis set.}\label{fig:MO}
\end{figure}

The choice of active space is critical for excited state calculations.
For the ground state and low-lying electronic states, use of the valence MOs as the active space is generally sufficient, but a more careful selection must be made for high-energy and Rydberg states.
As mentioned previously, the experimental spectroscopy\cite{Herzberg1969a} and earlier theoretical calculations\cite{Bruna2001} have shown that the $F$ state is a Rydberg state with the configuration $\sigma_u^2 \pi_u^3 3s\,[1 ^2\Pi_u,3s]$, suggesting that inclusion of the 4$a_g$ orbital into the active space is necessary for accurate treatment of static electron correlation in the $F$ state (see Figure~\ref{fig:MO}).
After exploratory calculations, we also added the 5$a_g$, 6$a_g$, and 7$a_g$ MOs into the active space to achieve smooth PECs over the entire range of internuclear distances.
Our final CAS (8,12) active space contains 12 total MOs (7,1,1,0,3,1,1,0) and 8 valence electrons; the lowest two core MOs (1,0,0,0,1,0,0,0) are closed and doubly-occupied.
Rydberg states with 3s, 3d, and potentially 4s Rydberg orbitals are able to be well calculated in our study, while any Rydberg states with a 3p Rydberg orbital are absent in our results.
Though it is possible for states with 3p Rydberg orbitals to be $^1\Sigma_u^+$ or $^1\Pi_u$ states, near the equilibrium geometry of the $F\,^1\Pi_u$ state they are unlikely to contribute to the electronic character owing to their higher energies.

The DW-SA-CASSCF procedure was used to optimize the orbital shapes and establish the reference functions for the subsequent MRCI+Q calculations.
SA-CASSCF involves optimizing orbitals by minimizing the average energy of a set of electronic states with a specified spin multiplicity and symmetry, and has been found to yield smooth PECs for both the ground electronic state and excited states at the same time.
For internuclear distances below 3.2~\AA, (15,10,10,7,5,6,6,4), (8,13,13,11,14,9,9,12), and (3,1,1,4,1,3,3,3) singlet, triplet, and quintet states were averaged, respectively, and for larger internuclear distances, (11,5,5,5,3,5,5,5), (4,7,7,6,7,7,7,6), and (4,2,2,1,1,2,2,1) states were averaged.
The set of states for internuclear distance larger than 3.0~\AA\ correspond to the 6 atomic limits $^3P+{^3P}$ (defined as $E$=0), $^3P+{^1D}$, $^1D+{^1D}$, $^3P+{^1S}$, $^1D+{^1S}$, $^3P+{^5S}$ ($E$=0.154~Hartree). The diabatic F state converges to the $^3P+{^3P^o}$ limit theoretically, which lies at $E=$0.275~Hartree.
The selection of states at shorter internuclear distance depends on the SA-MCSCF energy at $R=$1.25~\AA.
All states with an energy within 0.50 Hartree (91.2~nm, 13.6~eV) of the ground $X$ state are included.
For states with $^1\Sigma_g^+$, $^1\Pi_u$, $^3\Pi_u$, $^3\Sigma_u^+$ and $^3\Sigma_u^-$ symmetry, the thresholds are set to 0.60~Hartree.
The calculations around 3.2\,\AA\ using both sets of averaged states differ by only $\sim$10~cm$^{-1}$.
After including such a large number of states in the SA-CASSCF calculation, the ground electronic state may not be well-optimized if all states have equal weights in the optimization process.
In the dynamically weighted state-averaged method\cite{Deskevich2004}, the weight for each desired state $W(x)$ varies dynamically based on the formula:
\begin{equation}
    W(x) = \text{sech}^2(\beta(E_x-E_0)),
\end{equation}
where $E_x$ and $E_0$ are the energy of each desired state and ground state, and $\beta$ is a parameter to control how fast the weight decreases as the energy increases.
DW-SA-CASSCF has been applied in several quantum chemical calculations involving excited states \cite{Dawes2010,Samanta2014}.
Here, we choose $\beta = 3.0$\,Hartree$^{-1}$; an excited state at 75,000 cm$^{-1}$ therefore has a weight of about 40\% compared to the ground state.
We did not find a significant difference from using dynamic weighting versus averaged weighting at the equilibrium configuration, as both methods yield similar sets of MOs and reference states for the following MRCI calculations.
At the equilibrium geometry, the relative MRCI+Q energy differences between the two are of the order of $10^{-4}$ to $10^{-3}$ Hartree.
The DW-SA-CASSCF MOs at an internuclear distance of 1.25~\AA\ are shown in Figure \ref{fig:MO}.
As the internuclear distance increases, the shapes and energy ordering of the MOs change significantly.

The configuration state functions calculated in the DW-SA-CASSCF procedure are used to generate the reference space in the following MRCI+Q calculations.
For the calculation of the $^1\Pi_u$ states, there are 4,060 configuration state functions (CSFs) in our reference space.
From the reference space, a total of 4,563,905 contracted CSFs formed from 91,843,656 uncontracted CSFs are included in the MRCI calculations.
PECs are computed with the Davidson correction added, and the transition dipole moments (TDMs) for allowed transitions from the ground $X$ state are evaluated from the MRCI wavefunctions.
Additionally, to study the perturbations and predissociation of the $F\,^1\Pi_u$ state, spin-orbit couplings (SOCs) and non-adiabatic coupling matrix elements (NACMEs) involving the $F$ states are also calculated using the MRCI wavefunctions.
The full Breit-Pauli operator is used to calculate the SOC matrix elements between internal configurations and a mean-field one-electron Fock operator is applied to calculate the contribution of external configurations.
The NACMEs are calculated by finite differences of the MRCI wavefunctions at $\Delta R=+0.001$\,\AA.

To further explore the Rydberg nature of the $F$ state, we calculated PECs for the two lowest $^2\Pi_u$ electronic states of \ce{C2+} using a valence CAS(7,8) active space and the same basis set.
Finally, for the ground state and most low-lying excited states, spectroscopic constants, including $T_e$, $\omega_e$, $\omega_ex_e$, $B_e$, $D_e$ and $\alpha_e$, were calculated by fitting the rovibrational energy levels derived from a numerical evaluation of the one-dimensional Schr\"{o}dinger equation using the DUO package\cite{Yurchenko2016}.
The dissociation limits $D_e$ are calculated as the energy difference of corresponding atomic limits and the potential well minimum.

\subsection{Photodissociation cross sections and photodissociation rates}\label{subsec:C2_theo_theory_pd_cross_section}

We apply the coupled-channel Schr\"{o}dinger equation (CSE) technique to study the predissociation mechanics of \ce{C2} states, focusing on the $F\,^1\Pi_u$ state.
This method was adapted from scattering theory\cite{Mies1980} and has been detailed in previous studies\cite{vanDishoeck1984, Heays2010}.
The CSE method has been used to study the photodissociation of many diatomic molecules, including \ce{OH}\cite{vanDishoeck1984}, \ce{O2}\cite{Gibson1996,Lewis2001}, \ce{N2}\cite{Heays2015}, and \ce{S2}\cite{Lewis2018}.
In those studies, coupled-channel models of states contributing to predissociation are built using a basis of diabatic states.
Then, least-squares fitting programs are used to optimize the model parameters, which include potential energy curves, transition dipole moments and couplings between states, to match the calculated cross sections to experimentally measured cross sections and linewidths.
We successfully employed this method, using \textit{ab initio} model parameters computed at the MRCI level to study the predissociation of \ce{CS} $^1\Sigma^+$ states\cite{Xu2019}.
In this study, the predissociation diabatic coupled channel model of \ce{C2} is built by including PECs of excited states of interest and the appropriate SOC values and non-adiabatic couplings among them.
Then the coupled-channel model is solved numerically with the python package PyDiatomic\cite{Gibson2016} to yield the coupled-channel wavefunctions for the excited states.
The total photodissociation cross sections can be calculated by combining the wavefunction for the uncoupled ground state, the coupled-channel wavefunction for the excited states, and the diabatic TDMs.
In the photodissociation cross section calculation, it is assumed that the photodissociation efficiency is essentially unity, which means all photoabsorption leads to photodissociation.
This assumption can be verified by comparing the predissociation lifetime derived from the calculated linewidth with the spontaneous emission lifetime.

The photodissociation rate of a molecule in a UV radiation field can be calculated as 
    \begin{equation}
    k = \int \sigma(\lambda) I(\lambda) d\lambda
    \end{equation}
where $\sigma(\lambda)$ is the photodissociation cross section including both direct photodissociation and predissociation and $I(\lambda)$ is the spectral photon flux density (photons s$^{-1}$ cm$^{-2}$ nm$^{-1}$) of the radiation field.
We compute the photodissociation rate of \ce{C2} from its ground ($X$) state with $(v^{\prime\prime}, J^{\prime\prime}) =(0,0)$ in the standard interstellar radiation field (ISRF)\cite{Draine1978} and several other radiation fields.

\section{Results\label{sec:C2_theo_results}}

\subsection{PECs}

\begin{figure*}
 \centering
 \begin{tikzpicture}
  \node[anchor=south west] at (0,10.5){\includegraphics[width=12cm]{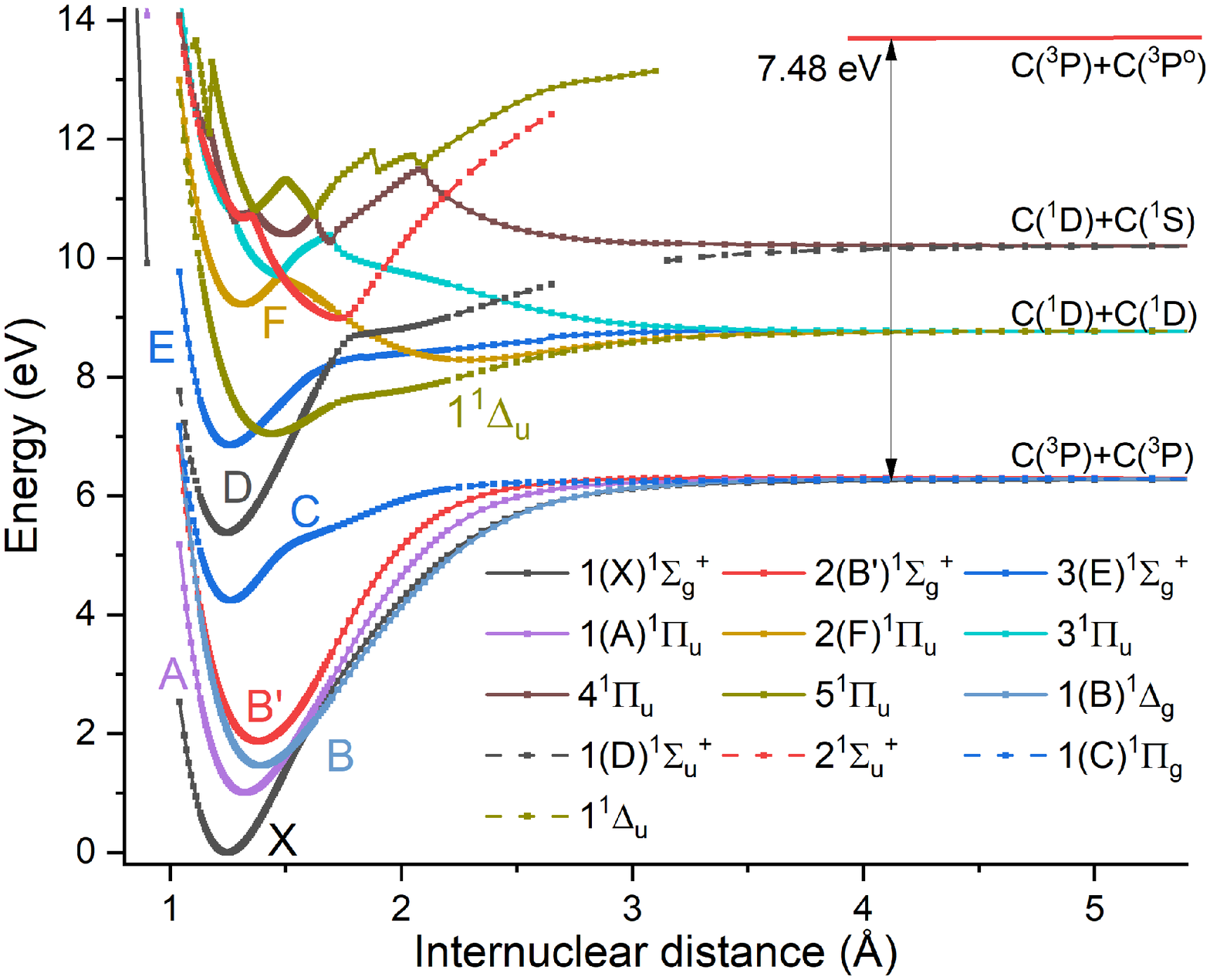}};
  \node[anchor=south west] at (0,0){\includegraphics[width=12cm]{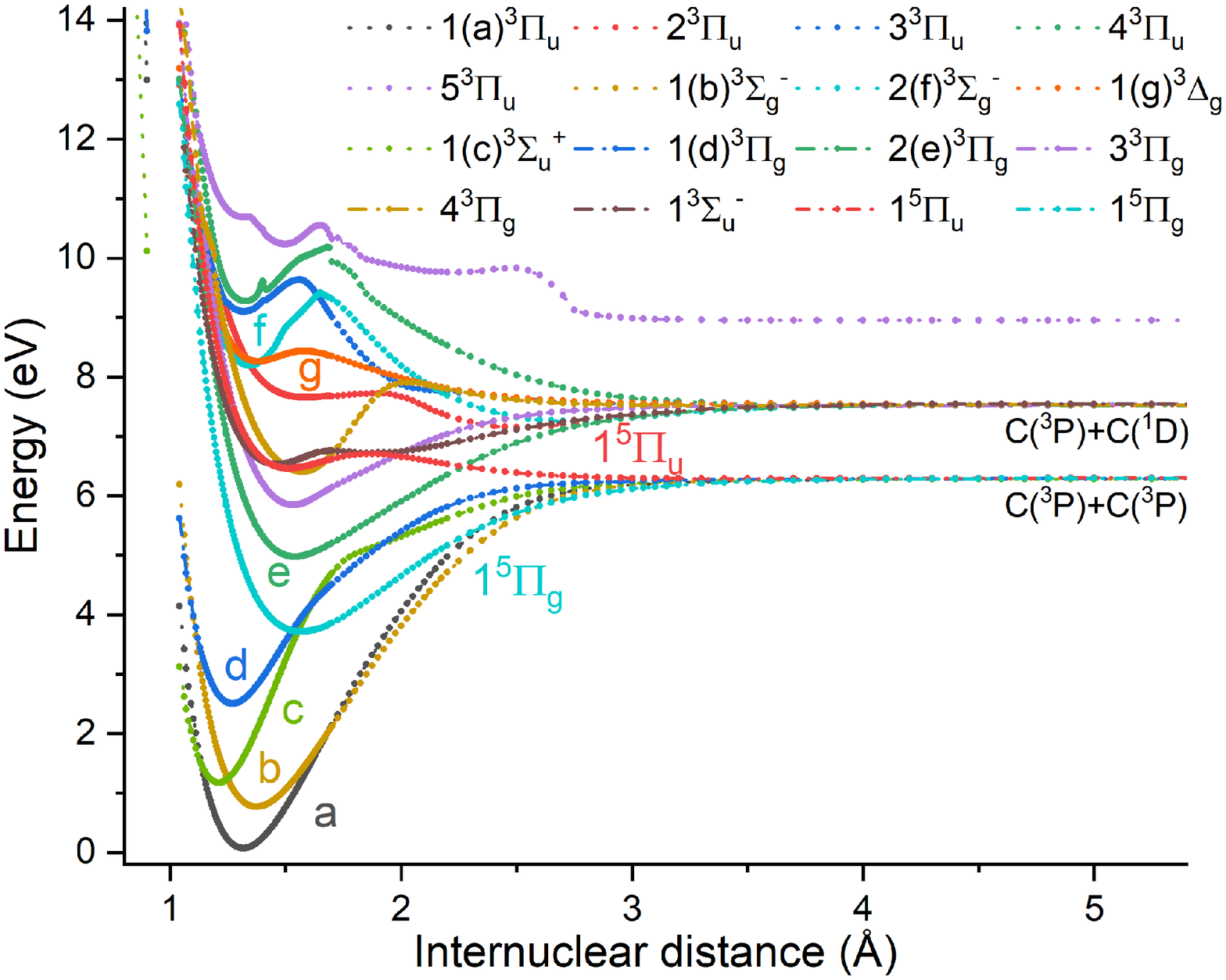}};
 \end{tikzpicture}
 \caption{PECs (eV) as a function of internuclear distance (\AA) for selected \ce{C2} (top) singlet and (bottom) triplet and quintet states, calculated at DW-SA-CASSCF/MRCI+Q with CAS (8,12) active space and aug-cc-pV5Z+2s2p basis set. All experimentally studied states and several other states related to predissociation through the $F\,^1\Pi_u$ state are included.}\label{fig:PECs}
\end{figure*}

\begin{figure*}
 \centering
 \begin{tikzpicture}  
  \node[anchor=south west] at (0,13){\includegraphics[width=8.5cm]{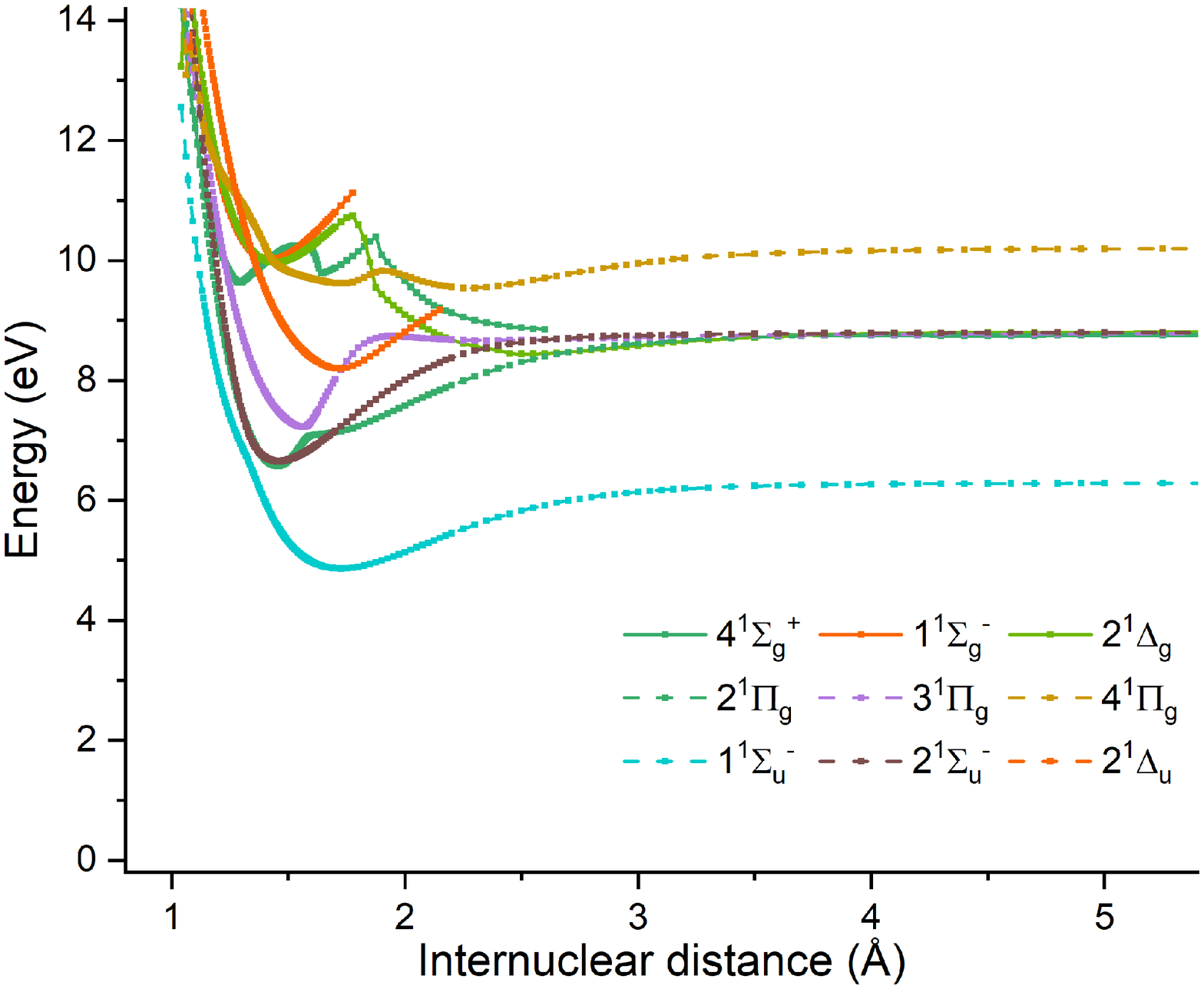}};
  \node[anchor=south west] at (0,6.5){\includegraphics[width=8.5cm]{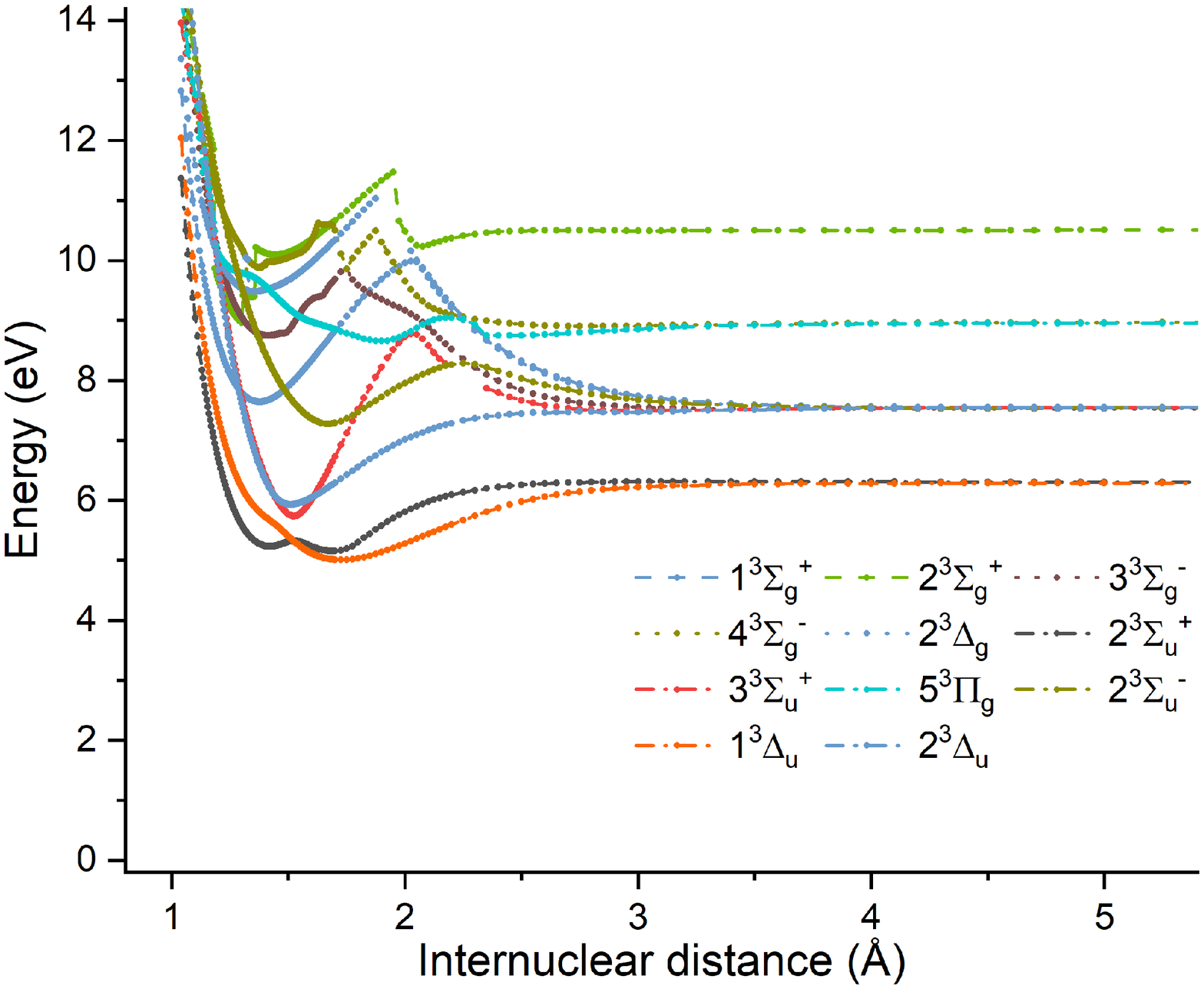}};
  \node[anchor=south west] at (0,0){\includegraphics[width=8.5cm]{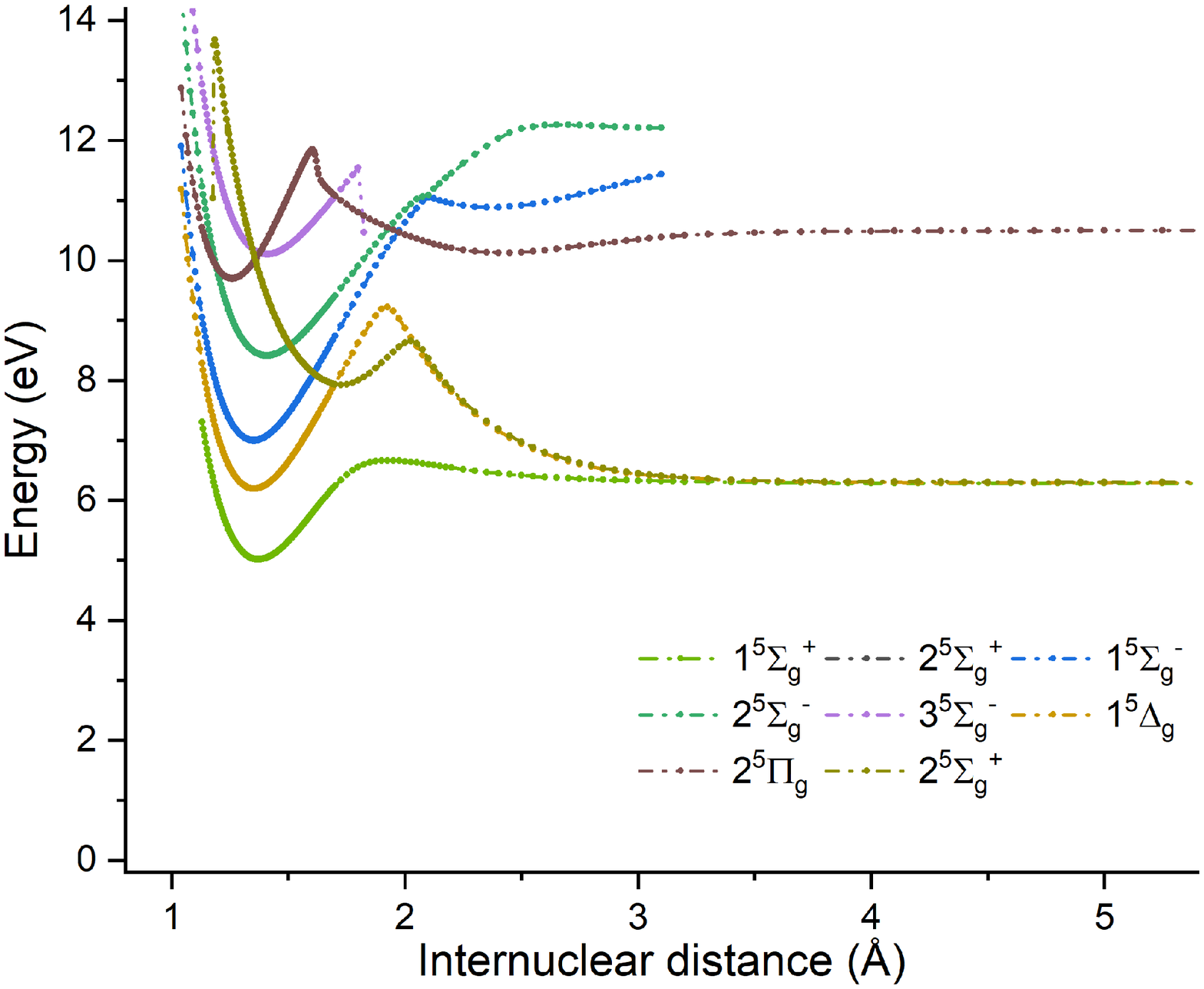}};
 \end{tikzpicture}
 \caption{PECs (eV) as a function of internuclear distance (\AA) for \ce{C2} (top) singlet, (middle) triplet, and (bottom) quintet states which are not directly related to predissociation though the $F\,^1\Pi_u$ state.}\label{fig:PECs_sup}
\end{figure*}

We have successfully calculated the PECs of 56 states in total, some of which only have PECs available in a specific range of internuclear distances.
The PECs of states which will be discussed in depth in this work are shown in Figure~\ref{fig:PECs}, while the PECs of other singlet, triplet, and quintet states are shown in Figure~\ref{fig:PECs_sup}.
To estimate the accuracy of our calculations, a comparison between the calculated spectroscopic constants and a selection of experimental values is shown in Table~\ref{tab:spectroscopic_constant}.
Experimental spectra show abundant perturbations among excited states of \ce{C2}, so the polynomial fitted basic spectroscopic constants cannot fully reproduce the experimental spectra.
The spectroscopic constants shown here only provide a description of the shapes of calculated adiabatic potential energy curves.
We will first discuss experimentally known states, followed by a brief discussion of other states.
Quintet states were found to be unimportant for the photodissociation of \ce{C2} in space and will not be considered further here.

The 11 low-lying states of \ce{C2} primarily involve the molecular orbitals $2\sigma_u$, $1\pi_u$, and $3\sigma_g$.
Based on the orbital energies calculated at DW-SA-CASSCF at $R=1.25$\,\AA, the $3\sigma_g$ orbital is only 0.63\,eV higher than the $1\pi_u$ orbitals, while $1\pi_u$ is about 3.52\,eV higher than the $2\sigma_u$ orbital.
Keeping the core $1\sigma_g$ and $1\sigma_u$ orbitals, and the first valence orbital $2\sigma_g$ doubly-occupied, these low lying electronic states arise from configurations with 6 electrons distributed among the $2\sigma_u$, $1\pi_u$, and $3\sigma_g$ orbitals.
The 11 states coming from these 6 configurations are listed in Table~\ref{tab:conf_low}.
It is well known that the ground $X$ state of \ce{C2} has a multi-reference nature.
The $2\sigma_u^2\, 1\pi_u^4\, 3\sigma_g^0$ in the table is only the dominant configuration around the potential minimum.
The other states near the photodissociation threshold, including $1\,^1\Delta_u$, $e\,^3\Pi_g$, $3\,^3\Pi_g$ and $4\,^3\Pi_g$, have equilibrium distances much longer than the ground $X$ state; thus the dominant configurations are more complicated.
Few studies have been done on the remaining three excited states.
The $F\,^1\Pi_u$ state has been suggested as a Rydberg state corresponding to $1\pi_u \rightarrow 3s$ in several previous studies\cite{Pouilly1983,Bruna2001}.
Bruna and Grein found that $f\,^3\Sigma_g^-$ is a mixed valence-Rydberg state while $g\,^3\Delta_g$ is a valence state.

\begin{table}
  \centering
  \footnotesize
\caption{\label{tab:spectroscopic_constant}Spectroscopic constants for several low-lying and highly excited states in cm$^{-1}$ of \ce{C2}, along with equilibrium bond lengths in \AA\ and dissociation limits in eV.}
\begin{tabularx}{0.9\textwidth}{@{}rrrrrrrrrrr@{}}
    \hline \hline
                         & \multicolumn{1}{c}{Method} & \multicolumn{1}{c}{$T_e$} & \multicolumn{1}{c}{$\omega_e$} & \multicolumn{1}{c}{$\omega_e x_e$} & \multicolumn{1}{c}{$\omega_e y_e$} & \multicolumn{1}{c}{$B_e$} & \multicolumn{1}{c}{$D_e$} & \multicolumn{1}{c}{$\alpha_e$} & \multicolumn{1}{c}{$r_e$} & \multicolumn{1}{c}{$D_e$} \\ 
                         & & (cm$^{-1}$) & (cm$^{-1}$) & (cm$^{-1}$) & (cm$^{-1}$) & (cm$^{-1}$) & (10$^{-6}$cm$^{-1}$) & (cm$^{-1}$) & (\AA) & (eV) \\
                         \hline

		         $X\,^1\Sigma_g^+$ &   This work                  &           0 &    1844.178 &      12.553 &      $-$0.358 &              1.81281 &       6.944 &  0.02205 & 1.2449 & 6.2707 \\
		                           &        Expt$^a$              &           0 &    1855.035 &      13.570 &      $-$0.127 &              1.82005 &       6.972 &  0.01790 &        &        \\
		              $A\,^1\Pi_u$ &   This work                  &    8115.177 &    1594.881 &      11.425 &       $-$0.06 &              1.60703 &       6.505 &  0.01762 & 1.3222 & 5.2873 \\
		                           &        Expt$^a$              &    8391.406 &    1608.217 &      12.078 &      $-$0.003 &              1.61660 &       6.505 &  0.01693 &        &        \\
		           $B\,^1\Delta_g$ &   This work                  &   11795.780 &    1398.345 &      11.477 &       0.015 &              1.45411 &       6.307 &  0.01682 & 1.3900 & 4.8285 \\
                               &        Expt$^b$              &   12082.343 &    1407.451 &      11.471 &       0.009 &              1.46367 &       6.306 &  0.01681 &                 \\
    $B^{\prime}\,^1\Sigma_g^+$ &   This work                  &   15083.953 &    1409.625 &    $-$2.306 &      $-$0.413 &              1.47361 &       6.263 &  0.01084 & 1.3808 & 4.4286 \\
		                           &        Expt$^b$              &   15410.33\invis{0}  &    1420.36\invis{0}  &      &  &              1.47967 &       6.785 &  0.00943 &                 \\
                  $C\,^1\Pi_g$ &   This work                  &   34221.241 &    1762.866 &       1.173 &   $-$7.89\invis{0} &         1.76747 &       7.092 &  0.05462 & 1.2608 & 2.0506 \\
		         $D\,^1\Sigma_u^+$ &   This work                  &   43416.653 &    1810.257 &      14.616 &      -0.005 &              1.81134 &       7.320 &  0.01949 & 1.2454 & 4.8294 \\
                               &        Expt$^c$              &   42315.83\invis{0} & 1829.905 & 14.089 &       0.001 &              1.83254 &             &  0.01909 &                 \\
		         $E\,^1\Sigma_g^+$ &   This work                  &   55297.535 &    1663.568 &      43.806 &       0.531 &              1.77829 &       8.439 &  0.06309 & 1.2570 & 1.9187 \\
                              &         Expt$^d$              &             & 1671.5\invis{00} & 40.02\invis{0} &     &              1.7930\invis{0} &     &  0.0421\invis{0}        & 1.25\invis{00} & \\
		           $1\,^1\Delta_u$ &   This work                  &   56842.617 &    1089.398 &   $-$14.306 &     $-$7.302 &              1.35977 &       7.964 &  0.04574 & 1.4374 & 1.7329 \\ \hline
		              $a\,^3\Pi_u$ &   This work                  &     581.232 &    1629.966 &      11.693 &       0.001 &              1.62586 &       6.425 &  0.01726 & 1.3146 & 6.2253 \\
		                           &        Expt$^a$              &     720.008 &    1641.326 &      11.649 &     $-$0.002 &              1.63231 &       6.448 &  0.01654 &        &        \\
		         $b\,^3\Sigma_g^-$ &   This work                  &    6211.833 &    1459.059 &      10.865 &     $-$0.018 &              1.49223 &       6.205 &  0.01674 & 1.3722 & 5.5225 \\
		                           &        Expt$^a$              &    6439.083 &    1470.365 &      11.135 &       0.010 &              1.49866 &       6.221 &  0.01629 &        &        \\
		         $c\,^3\Sigma_u^+$ &   This work                  &    9466.486 &    2048.183 &      12.414 &     $-$0.328 &              1.91621 &       6.736 &  0.02005 & 1.2109 & 5.1102 \\
                               &        Expt$^e$              &    8662.925$^f$ &2061.940 &      14.836 &             &              1.9319\invis{0}  &    &  0.01855 &                 \\  
		              $d\,^3\Pi_g$ &   This work                  &   20207.845 &    1766.732 &      12.560 &    $-$1.117 &              1.74613 &       6.777 &  0.02988 & 1.2685 & 3.7920 \\
		              $e\,^3\Pi_g$ &   This work                  &   40142.394 &    1100.606 &      29.400 &       0.771 &              1.18297 &       5.737 &  0.02163 & 1.5411 & 2.5459 \\
		              $3\,^3\Pi_g$ &   This work                  &   47149.632 &    1319.834 &      79.582 &       3.777 &              1.20085 &       4.665 &  0.02861 & 1.5296 & 1.7079 \\
		              $4\,^3\Pi_g$ &   This work                  &   51691.940 &    1226.038 &   $-$12.212 &     $-$1.721 &              1.13718 &       3.863 &  0.07995 & 1.5718 & 1.1315 \\ \hline
		              $1\,^5\Pi_g$ &   This work                  &   29981.496 &     963.787 &       6.245 &    $-$0.196 &              1.14275 &       6.414 &  0.01592 & 1.5680 & 2.5875 \\ \hline\hline
\end{tabularx}
 \\
  a. Ref.~\cite{Chen2015a} \quad 
  b. Ref.~\cite{Chen2016} \quad 
  c. Ref.~\cite{Krechkivska2018} \quad 
  d. Ref.~\cite{Freymark1950} \quad
  e. Ref.~\cite{Joester2007} \quad 
  f. $V_{00}$ from $a\,^3\Pi_u$ state
\end{table}

\begin{table}[]
\caption{\label{tab:conf_low}The configurations with $2\sigma_u\, 1\pi_u\, 3\sigma_g$ orbitals and their corresponding electronic states}
\begin{tabularx}{0.5\textwidth}{@{}p{10em}l@{}}
    \hline \hline
    configuration & electronic state \\ \hline
    $2\sigma_u^2\, 1\pi_u^4\, 3\sigma_g^0$ & $X\,^1\Sigma_g^+$ \\
    $2\sigma_u^2\, 1\pi_u^3\, 3\sigma_g^1$ & $A\,^1\Pi_u$, $a\,^3\Pi_u$ \\
    $2\sigma_u^2\, 1\pi_u^2\, 3\sigma_g^2$ & $B\,^1\Delta_g$, $B^{\prime}\,^1\Sigma_g^+$, $b\,^3\Sigma_g^-$ \\
    $2\sigma_u^1\, 1\pi_u^4\, 3\sigma_g^1$ & $D\,^1\Sigma_u^+$, $c\,^3\Sigma_u^+$ \\
    $2\sigma_u^1\, 1\pi_u^3\, 3\sigma_g^2$ & $C\,^1\Pi_g$, $d\,^3\Pi_g$ \\
    $2\sigma_u^0\, 1\pi_u^4\, 3\sigma_g^2$ & $E\,^1\Sigma_g^+$ \\
    \hline \hline
\end{tabularx}
\end{table}

\subsubsection{Low-lying states}

For the low-lying electronic states, thousands of high-resolution rovibronic lines have been recorded.
Using those transitions, Chen \textit{et al.} determined the energy difference between the ground $X\,^1\Sigma_g^+$ and $a\,^3\Pi_u$ states to be 720.008(2) cm$^{-1}$ and derived updated spectroscopic constants for $X\,^1\Sigma_g^+$, $A\,^1\Pi_u$, $a\,^3\Pi_u$ and $b\,^3\Sigma_g^-$ states\cite{Chen2015a}. 
Because our calculation averages many electronic states in the CASSCF procedure to treat high-energy states, it is reasonable that the accuracy for low-lying states is diminished relative to calculations focusing only on those states.
Nevertheless, our calculation still shows good agreement with the available experimental data.
The $X$ state is deeply bound with a dissociation limit $D_e$ 6.2707eV.
The calculated vibrational constant $\omega_e$ is 1844.178\,cm$^{-1}$, which is about 11\,cm$^{-1}$ smaller than the experimental value.
The $a\,^3\Pi_u$ state is only 581.232\,cm$^{-1}$ higher than the ground $X$ state in our calculation, which is about 140\,cm$^{-1}$ smaller than the experimental value derived\cite{Chen2015a}.
Despite the electronic energy difference, the harmonic vibrational constants of the $X\,^1\Sigma_g^+$ and $a\,^3\Pi_u$ states calculated by our method are quite close to the experimental values.

The singlet states $A\,^1\Pi_u$, $B\,^1\Delta_g$, and $B^{\prime}\,^1\Sigma_g^+$ have deep potential wells and converge to the $^3P+{^3P}$ atomic limit.
Compared with their recently updated spectroscopic constants\cite{Chen2015a,Chen2016}, the $T_e$ values of these three states have been underestimated by our calculation by $\sim$300-500\,cm$^{-1}$, while the vibrational constants are similar ($|\Delta\omega_e|<15$\,cm$^{-1}$).
Another three singlet states $C\,^1\Pi_g$, $D\,^1\Sigma_u^+$, and $E\,^1\Sigma_g^+$ involve exciting electrons from the $2\sigma_u$ orbital to the $1\pi_u$ and $3\sigma_g$ orbitals. 
All have a potential minimum near $R=1.25$\,\AA, similar to the ground $X\,^1\Sigma_g^+$ state, and much smaller than those of $A$,$B$, and $B^{\prime}$ states, which are beyond $1.3$\,\AA. 
Because the $2\sigma_u$ orbital is an anti-bonding orbital, removing electrons would not be expected to decrease the bond order nor weaken the bond strength significantly.
For similar reasons, the vibrational constants of these three states are significantly larger than those of $A$, $B$, and $B^{\prime}$ states.
The $C\,^1\Pi_g$ state has a notable avoided crossing with $2\,^1\Pi_g$, which has been well described by a diabatic valence-hole model.~\cite{Jiang2022}
The adiabatic $D\,^1\Sigma_u^+$ state has an avoided crossing with the adiabatic $2\,^1\Sigma_u^+$ state at $R=1.75$\,\AA.
From the shape of the two adiabatic curves, the diabatic $2\,^1\Sigma_u^+$ state  in this region has a potential well around $R=1.90$\,\AA, which is very close to the avoided crossing point.
The large difference in equilibrium bond length between this adiabatic state and the ground $X\,^1\Sigma_g^+$ state indicates the Franck-Condon factors for the vibronic transitions between them would likely be too low to play an important role in the photodissociation of \ce{C2}.

The triplet states $b\,^3\Sigma_g^-$, $c\,^3\Sigma_u^+$, and $d\,^3\Pi_g$ have lower electronic energies compared with singlet states with similar configurations but different multiplicities.
For example, $d\,^3\Pi_g$ is about 23000\,cm$^{-1}$ lower than $C\,^1\Pi_g$, but they have similar $r_e$ and $\omega_e$ values with differences $\Delta r_e = 0.008$\,\AA\ and $\Delta \omega_e = 4$\,cm$^{-1}$.

Compared with experimental values, our theoretical $\omega_e$ values are consistently underestimated by about 10\,cm$^{-1}$ or more.
Including core-valence correlation has been shown to reduce the differences, as demonstrated by a previous theoretical study on low lying states of \ce{C2}.~\cite{Kokkin2007}

\subsubsection{$^1\Pi_u$ states}

We will focus on $^1\Pi_u$ states here because these states are directly related to the photodissociation of \ce{C2}.
The $A\,^1\Pi_u$ state lies only about 1 eV above the ground $X$ state.
Our calculated $T_e$ is 8115.177 cm$^{-1}$, which is a bit smaller than the experimental value 8391.406 cm$^{-1}$\cite{Chen2015a}.
Our calculated vibrational constant 1594.881 cm$^{-1}$ shows good agreement with the experimental value of 1608.217 cm$^{-1}$.
The calculated $2\,^1\Pi_u$ state has a double well structure, with an inner well located at 1.32\,\AA\ and an outer shallow well located at 2.25\,\AA.
The dominant configuration of the $2\,^1\Pi_u$ state at 1.32\AA\ is $2\sigma_g^2\, 2\sigma_u^2\, 1\pi_u^3\, 4\sigma_g$.
This inner potential well of $2\,^1\Pi_u$ corresponds to the experimental $F\,^1\Pi_u$ state which was first discovered by Herzberg, Lagerqvist,
and Malmberg\cite{Herzberg1969a}.
Our calculation agrees that the $F\,^1\Pi_u$ state in this region is a Rydberg state with configuration $\sigma_u^2\pi_u^33s$ or [$^2\Pi_u,3s$].
Then, the $2\,^1\Pi_u$ state has an avoided crossing with the $3\,^1\Pi_u$ state at 1.47 A.
The electronic structure of the outer potential well is more complicated and unable to be represented by just one primary configuration.
A similar double-well structure has also been reported\cite{Bruna2001}.
Hereafter, the $2\,^1\Pi_u$ and $F\,^1\Pi_u$ labels are used interchangeably to refer to the Herzberg $F$ state.
The PECs of $^1\Pi_u$ states at energies higher than the $F$ state contain frequent avoided crossings all over the internuclear distance range, with segments corresponding to different bound or non-bound states.
From the PECs, at least another two bound states can be recognized.
The first bound state contains these main segments built from $3\,^1\Pi_u$ at $R<$1.3\,\AA, $4\,^1\Pi_u$ from 1.3 to 1.4\,\AA, and $5\,^1\Pi_u$ from 1.4 to 1.5\,\AA.
The dominant configuration of this state is $2\sigma_g^2\, 2\sigma_u^2\, 1\pi_u^3\, 5\sigma_g$.
The other bound $^1\Pi_u$ state has the potential well of the $4\,^1\Pi_u$ state at 1.5\,\AA.

\subsubsection{Other triplet states}

$^3\Pi_u$ states are also important in \ce{C2} photodissociation because they have spin-orbit couplings with $^1\Pi_u$ states.
No $^3\Pi_u$ other than the $a\,^3\Pi_u$ state have been studied by experiments so far.
The PECs of the 2-5\,$^3\Pi_u$ states were previously calculated and discussed\cite{Bruna2001}.
Our calculation shows some substantial differences.
In our calculation, the $2\,^3\Pi_u$ state is repulsive with avoided crossings with $3\,^3\Pi_u$ at $R=1.235$ and 1.98\,\AA.
The latter avoided crossing has a energy gap of 0.2\,eV, indicating a strong non-adiabatic coupling.
An almost flat shape is observed in the PEC from 1.5-2.0\,\AA. 
Both the $3\,^3\Pi_u$ and $4\,^3\Pi_u$ states have a deep potential well at $R=1.32$\,\AA\ and have a barrier at larger distance. 
The $4\,^3\Pi_u$ state has obvious discontinuities at $R=1.40$ and 1.69\,\AA, indicating some potential avoided crossings are not calculated correctly in our study.
Those corresponding states may not be well described by our active space.
For the same reason, The $5\,^3\Pi_u$ state calculated and shown here is likely inaccurate.

\subsubsection{Other singlet states}

The calculated PEC of the $E^1\Sigma_g^+$ state has a well defined Morse potential shape with an equilibrium distance 1.26\AA.
Although the $T_e$ of the $E^1\Sigma_g^+$ state is about 55000\,cm$^{-1}$ higher than the ground state, no avoided crossings are observed since the fourth $^1\Sigma_g^+$ state is at least 20000\,cm$^{-1}$ higher still.
This state has been detected through the $E^1\Sigma_g^+ - A\,^1\Pi_u$ bands twice previously\cite{Freymark1950,Sorkhabi1997a}.
The calculated $E-A$ (0-0) transition energy is 47216.70\,cm$^{-1}$, which is about 548\,cm$^{-1}$ above the experimental value of 46668.3\,cm$^{-1}$\cite{Freymark1950}.
The calculated vibrational constant is only 8\,cm$^{-1}$ smaller than the experimental value.

The last state shown in Figure~\ref{fig:PECs} is $1\,^1\Delta_u$, which has been only studied experimentally by Goodwin \text{et al.}\cite{Goodwin1988, Goodwin1989}.
The calculated PEC of the $1\,^1\Delta_u$ state has a potential minimum at $1.436$\,\AA, which matches the experimental value exactly.
The calculated $T_e$ in this work is about 878\,cm$^{-1}$ higher than the experimental value of 57720\,cm$^{-1}$, while the $\omega_e$ in our work is about 60\,cm$^{-1}$ smaller than the experimental value of 1150\,cm$^{-1}$.
In the PEC, a slight bending is observed around $1.75$\,\AA, indicating a strong adiabatic interaction with the $2\,^1\Delta_u$ state (shown in Figure~{\ref{fig:PECs_sup}}) with a coupling estimated as 0.3\,eV.
By constructing a diabatic $1\,^1\Delta_u$ state from the adiabatic PECs, the resultant $\omega_e$ value would be expected to lie much closer to experimental one.

\subsection{Important electronic transitions and their TDMs}

\begin{figure*}
  \centering
  \begin{tikzpicture}
   \node[anchor=south west] at (0,0){\includegraphics[width=8cm]{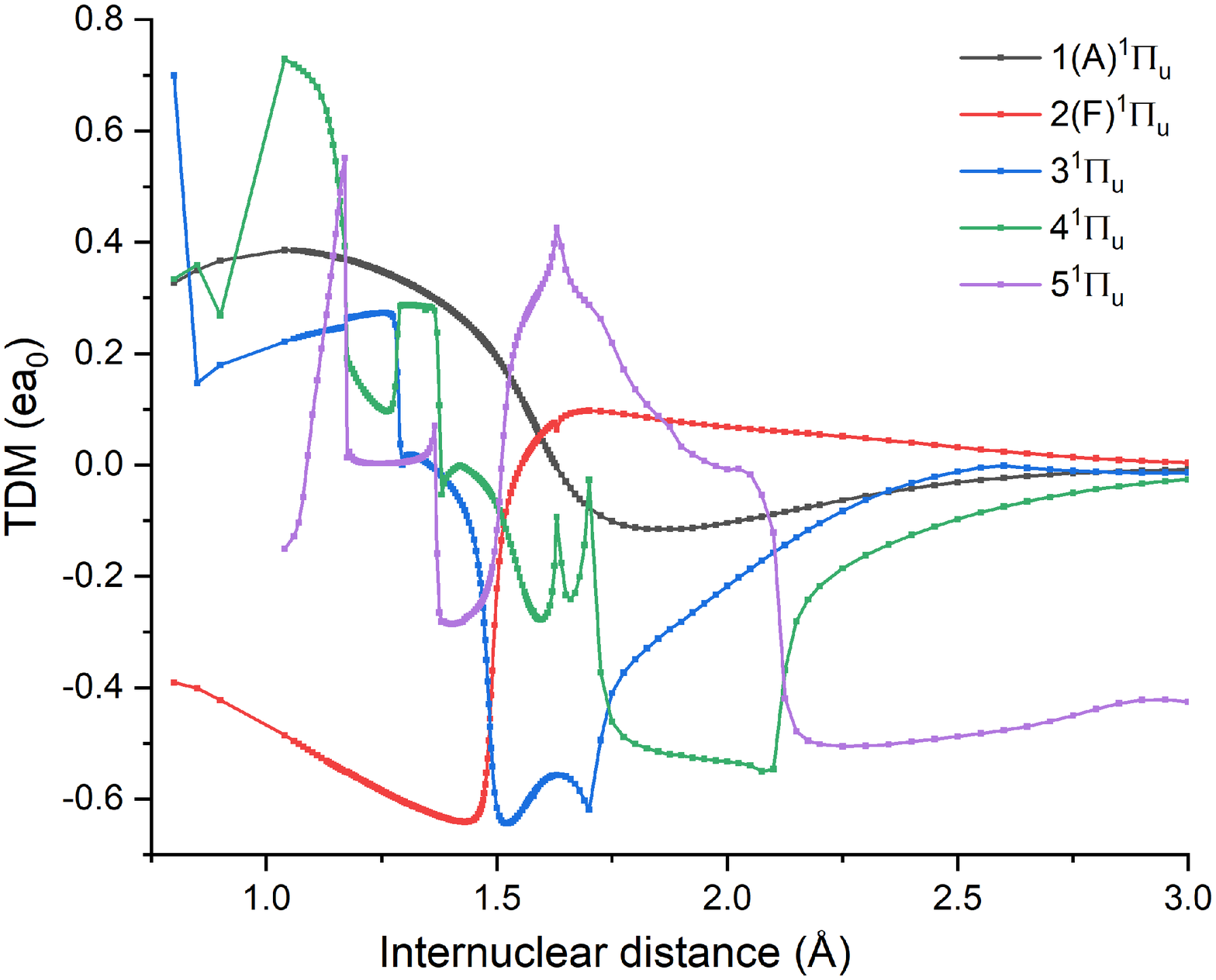}};
   \node[anchor=south west] at (8,0){\includegraphics[width=8cm]{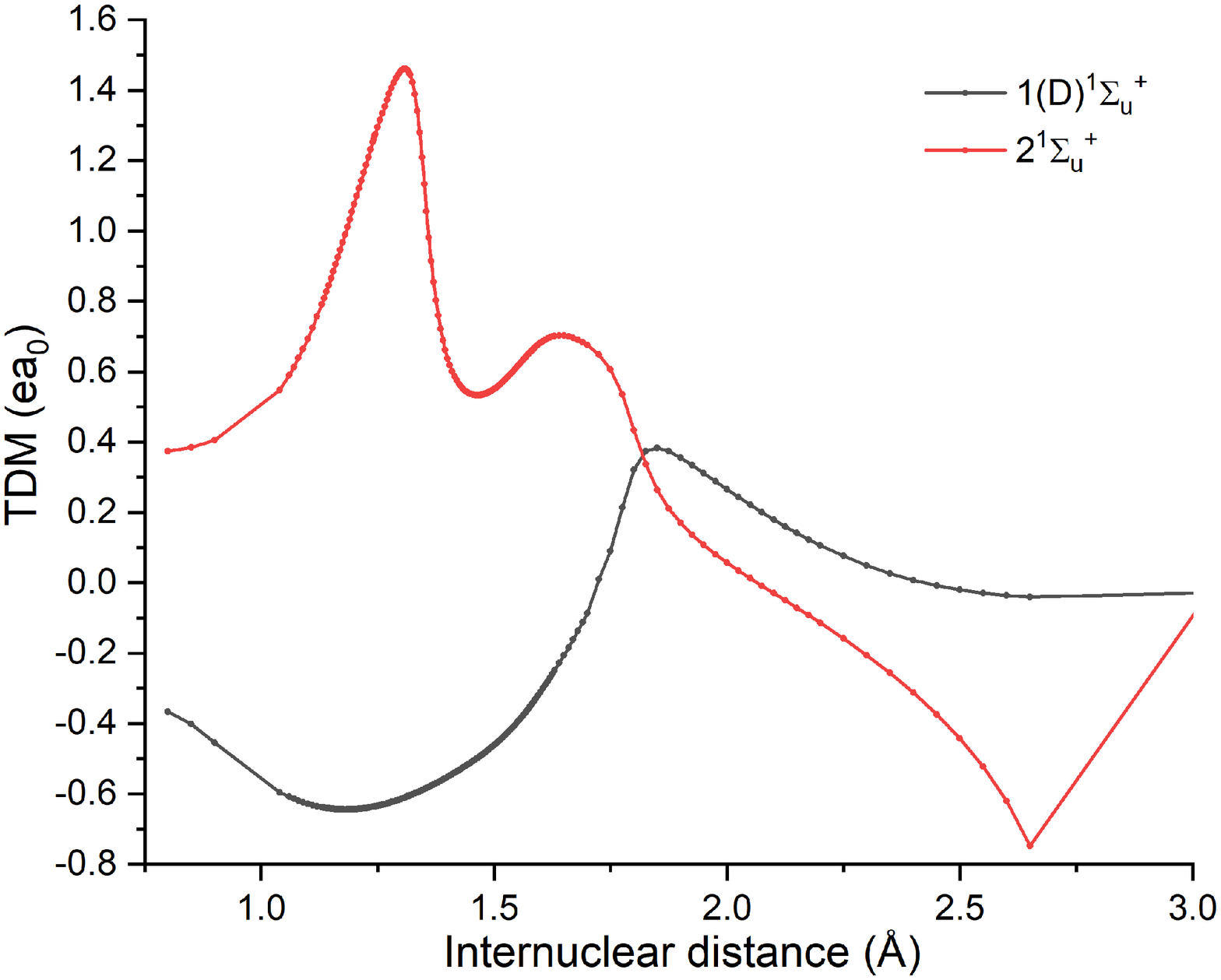}};
  \end{tikzpicture}
  \caption{TDMs for transitions of \ce{C2} from the $X\,^1\Sigma_g^+$ state to (left) $^1\Pi_u$ and (right) $^1\Sigma_u^+$ states in atomic units. Phases are manually corrected. }\label{fig:TDMs}
 \end{figure*}

\begin{figure}
  \centering
  \includegraphics[width=8cm]{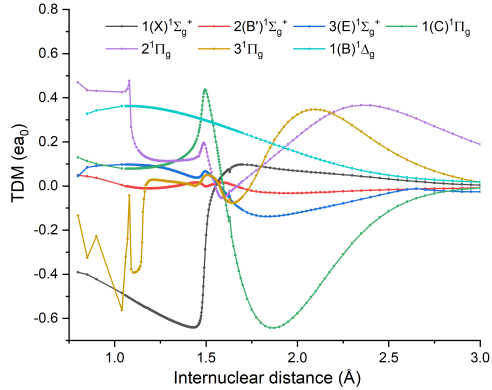}
  \caption{TDMs for transitions of \ce{C2} from the $F\,^1\Pi_u$ state to lower states. Phases are manually corrected.}\label{fig:TDMs_2}
 \end{figure}

The selection rules for electronic transitions between homonuclear diatomic atoms in Hund's case (a) and (b) are:
\begin{equation}
\Delta \Lambda = 0,\pm1; \quad \Delta S = 0; \quad + \nleftrightarrow -;\quad g \leftrightarrow u \label{equ:trans_selection}
\end{equation}
To study the photodissociation of \ce{C2}, we need to consider not only the absorption from the ground state to available excited states, but also the spontaneous emission from the excited states which may decrease their lifetime and compete with predissociation.
Among all states discussed so far, only those with $^1\Sigma_u^+$ and $^1\Pi_u$ symmetry can be directly excited from the ground $X\,^1\Sigma_g^+$ state.
The $^1\Sigma_u^+$ excited states can relax to $^1\Pi_g$ and $^1\Sigma_g^+$ states by spontaneous emission, while the $^1\Pi_u$ states can relax to $^1\Sigma_g^+$, $^1\Sigma_g^-$, $^1\Pi_g$, and $^1\Delta_g$ states.
For example, the $F\,^1\Pi_u$ state is able to relax to $X\,^1\Sigma_g^+$, $B^{\prime}\,^1\Sigma_g^+$, $E\,^1\Sigma_g^+$, $C\,^1\Pi_g$, $2\,^1\Pi_g$, $3\,^1\Pi_g$, and $B\,^1\Delta_g$ states.
Calculated TDMs for transitions relevant for \ce{C2} photodissociation are shown in Figure~\ref{fig:TDMs} and Figure~\ref{fig:TDMs_2}.
All the TDMs shown here are phase corrected manually.
It is well known that TDMs calculated by MOLPRO are in random phases at different internuclear distances; thus a manual correction needs to be done to assign the correct phases for the TDMs.

\begin{figure*}
 \centering
  \includegraphics[width=15cm]{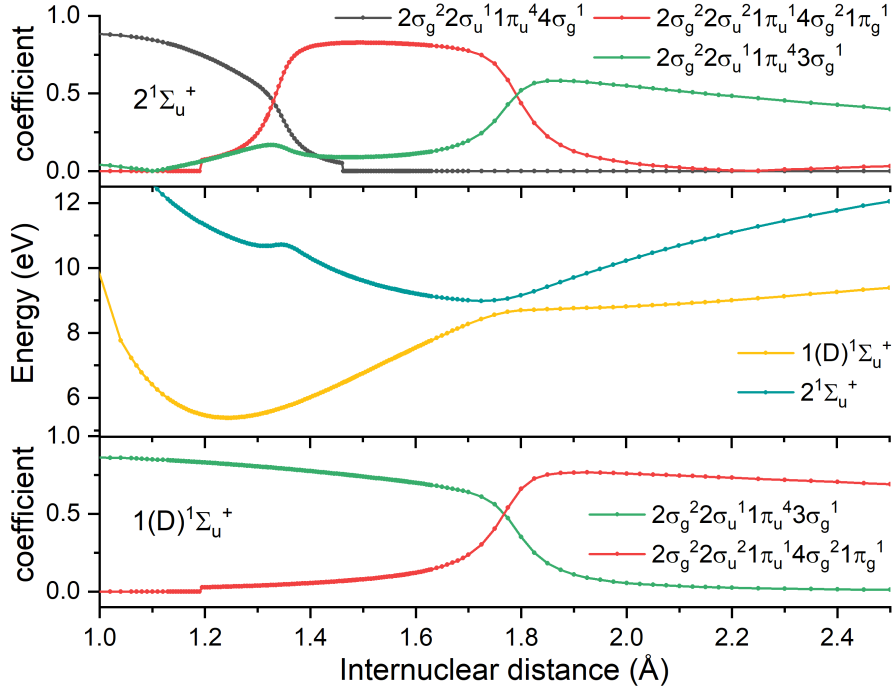}
 \caption{MRCI+Q PECs (middle) of the $2\,^1\Sigma_u^+$ and $D\,^1\Sigma_u^+$ states, with the coefficients of the most important electron configurations in the MRCI wavefunctions of $2\,^1\Sigma_u^+$ (top) and $D\,^1\Sigma_u^+$ (bottom). The same configuration is shown in same color between states.}\label{fig:conf_1Sigma_u_p}
\end{figure*}

As previously discussed, two $^1\Sigma_u^+$ states are calculated here, $D\,^1\Sigma_u^+$ and $2\,^1\Sigma_u^+$.
Although the  $v\geq 5$ vibrational levels of $D\,^1\Sigma_u^+$ state are calculated to lie above the photodissociation threshold, transitions from the ground $X\,^1\Sigma_g^+$ ($v=0$) state are expected to have small Franck-Condon factors, and the corresponding bands have never been detected experimentally.
The calculated $2\,^1\Sigma_u^+$ shows a double-well structure.
The potential well at $R=1.73$\,\AA\ arises from the avoided crossing with the $D\,^1\Sigma_u^+$ state.
The potential barrier at $R=1.35$\,\AA, which is only about 0.042\,eV above the potential well at $R=1.31$\,\AA, is from the avoided crossing with another higher energy state which is not included in the calculation.
As shown in Figure~\ref{fig:conf_1Sigma_u_p}, the dominant configuration of the adiabatic $2\,^1\Sigma_u^+$ state changes smoothly from Rydberg ($2\sigma_g^2\, 2\sigma_u^1\, 1\pi_u^4\, 4\sigma_g^1$) to valence ($2\sigma_g^2\, 2\sigma_u^2\, 1\pi_u^1\, 3\sigma_g^2\, 1\pi_g^1$) character though this avoided crossing.
The TDM for the $X\,^1\Sigma_g^+ - 2\,^1\Sigma_u^+$ transition shown in Figure~\ref{fig:TDMs} (right) is quite large around $R=1.25$\,\AA.
Therefore, it is expected that a strong absorption peak around $10.7$\,eV (116\,nm) can be observed, and could give rise to predissociation through the nonadiabatic coupling with lower diabatic $^1\Sigma_u^+$ states.
However, the experimental spectrum only covers 130-145\,nm\cite{Herzberg1969a} and the electronic states calculated previously only covers 7-10\,eV\cite{Bruna2001}.
This $2\,^1\Sigma_u^+$ state has not been reported by any previous studies, to the authors' knowledge.
Unfortunately, the PEC of the $3\,^1\Sigma_u^+$ state needed to construct a complete diabatic model of $2\,^1\Sigma_u^+$ is not calculated in this work, and thus a complete study on its absorption and dissociation is not carried out here.

Five $^1\Pi_u$ states are calculated in this work.
Besides $A\,^1\Pi_u$, all are above the photodissociation threshold.
As shown in Figure~\ref{fig:TDMs} (left), the TDM of the $F\,^1\Pi_u - X\,^1\Sigma_g^+$ transition is about 0.6\,$ea_0$ around $R=1.245$\,\AA, which is the equilibrium distance of the ground $X$ state.
The TDM from the ground $X$ state to the diabatic state which can be constructed from the $3\,^1\Pi_u$, $4\,^1\Pi_u$, and $5\,^1\Pi_u$ states is about 0.27\,$ea_0$ at $R=1.245$\,\AA.
The TDMs of transitions from $F\,^1\Pi_u$ to lower states are shown in Figure~\ref{fig:TDMs_2}.
Around $R=$1.245\,\AA, besides $F\,^1\Pi_u - X\,^1\Sigma_g^+$, only the $F\,^1\Pi_u - B\,^1\Delta_g$ transition has a modest TDM of about 0.35\,$ea_0$.
It can be expected that the spontaneous emissions from the $F\,^1\Pi_u$ state to other states is insignificant.

\section{Discussion\label{sec:C2_theo_diss}}

\subsection{Comparison With Previous Studies}

Abundant comparisons with experimental spectroscopic constants are already presented above.
To provide an better estimation of accuracy of the calculated PECs, we compare our results for the $D\,^1\Sigma_u^+$ state and several $^3\Pi_g$ states with previous studies.

As discussed in the Introduction, the $D\,^1\Sigma_u^+ - X\,^1\Sigma_g^+$ Mulliken bands have been studied in several experiments previously.
In short, the $v=0-4$ levels were recorded through the $D-X$ $\Delta v=0$ bands in the 20th century~\cite{Blunt1995} and more recently, the $D\,^1\Sigma_u^+$ $v=4-11$ levels were observed through $\Delta v=2$ bands~\cite{Krechkivska2018}.
A comparison between our results and these two previous experiments is shown in Table \ref{tab:comp_Mulliken}.
The energy difference from $v=0$ to $v=11$ is about 18000\,cm$^{-1}$.
Over this broad energy range, the difference between the calculated and the experimental $T_{v}$ values ($\Delta T_{v} = T_v^{expt}-T_{v}^{theory}$) ranges from $\Delta T_0$ = -171\,cm$^{-1}$ to $\Delta T_{11}$ = 103\,cm$^{-1}$.
Our calculated $B_v$ values are consistently 0.02\,cm$^{-1}$ smaller than experimental values, indicating that the calculated $r_e$ in Table~\ref{tab:spectroscopic_constant} is slightly too large.
The oscillator strength $f_{00}$ is calculated to be 0.05242, which is in good agreement with the experimental value 0.055$\pm$0.006, measured in 1969~\cite{Smith1969}.
For comparison, an MRCI/aug-cc-pV6Z calculation including relativistic corrections reported $f_{00}$ as 0.05346~\cite{Schmidt2007}.

\begin{table}[]
\caption{\label{tab:comp_Mulliken}Comparison between calculated $T_v$ and $B_v$ values of the $D\,^1\Sigma_u^+$ state with experimental values}
\begin{tabular}{@{}lllllllll@{}}
  \hline \hline
${v}$  & \multicolumn{2}{c}{$T_{v}$(cm$^{-1}$)}  & & \multicolumn{2}{c}{$T_{v} - T_{v-1}$(cm$^{-1}$)}  & & \multicolumn{2}{c}{$B_v$(cm$^{-1}$)}  \\
\cline{2-3} \cline{5-6} \cline{8-9}
  &   Expt            &  This work      & &   Expt       &   This work       & &  Expt                &    This work          \\ \hline
0$^a$  & 43227.33(40)  &43398.090 &&–        &          &&1.82322(15)      & 1.80641    \\
1$^a$  & 45028.87(33)  &45180.765 &&1801.54  & 1782.675 &&1.80370(39)      & 1.78673     \\
2$^a$  & 46802.45(23)  &46933.475 &&1773.58  & 1752.710 &&1.78390(06)      & 1.76779     \\
3$^a$  & 48547.83(25)  &48653.657 &&1745.38  & 1720.182 &&1.76470(50)      & 1.74869     \\
4$^a$  & 50258.27(15)  &50346.318 &&–        & 1692.661 &&1.74724(20)      & 1.72861     \\
4$^b$  & 50264.541(10) &          &&1716.711 &          &&1.74541(13)                     \\
5$^b$  & 51953.074(10) &52011.241 &&1688.533 & 1664.923 &&1.72495(18)      & 1.70826     \\
6$^b$  & 53612.649(19) &53645.902 &&1659.575 & 1634.661 &&1.70593(27)      & 1.68838     \\
7$^b$  & 55243.619(20) &55250.321 &&1630.97  & 1604.419 &&1.68588(29)      & 1.66793     \\
8$^b$  & 56845.597(66) &56826.277 &&1601.978 & 1575.956 &&1.66401(97)      & 1.64709     \\
9$^b$  & 58418.274(20) &58372.798 &&1572.677 & 1546.522 &&1.64491(25)      & 1.62633     \\
10$^b$ & 59961.329(14) &59888.199 &&1543.055 & 1515.400 &&1.62366(20)      & 1.60514     \\
11$^b$ & 61474.677(12) &61371.739 &&1513.348 & 1483.541 &&1.60355(17)      & 1.58330    \\
\hline \hline
\end{tabular}
 \\
  a. Ref.~\cite{Blunt1995} \quad 
  b. Ref.~\cite{Krechkivska2018} \quad 
\end{table}

\begin{figure*}
  \centering
   \includegraphics[width=10cm]{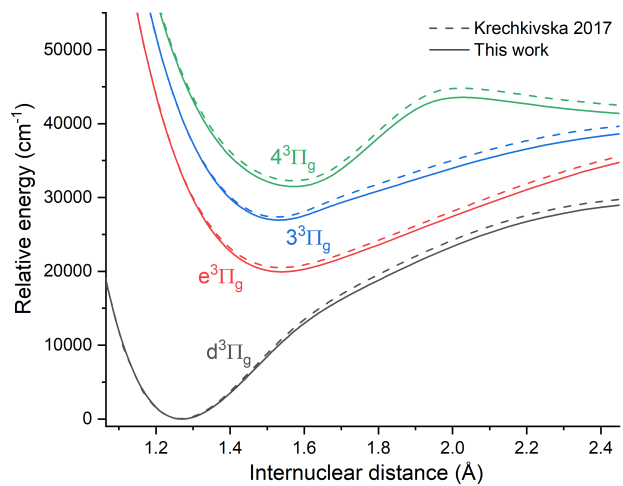}
  \caption{Comparison of $^3\Pi_g$ PECs calculated at CASSCF(8,8)-MRCI/aug-cc-pV6Z+Dav+CV+Rel\cite{Krechkivska2017} with the present DW-SA-CASSCF(8,12)-MRCI+Q/aug-cc-pV5Z+2s2p calculations (this work). The $T_e$ value of the $d\,^3\Pi_g$ state is set to 0 for both calculations.}\label{fig:comp_Krechkivska}
 \end{figure*}

The $e\,^3\Pi_g$ state was first studied through the $e-a$ transition and then through the $e-c$ transition\cite{Fox1937,Nakajima2009}.
The $V_{00}$ energy for the $e-a$ transition is 39296.5\,cm$^{-1}$, which is about 500\,cm$^{-1}$ lower than the value reported\cite{Fox1937}.
With the aid of \textit{ab initio} calculations, two new $^3\Pi_g$ states: $3\,^3\Pi_g$ and $4\,^3\Pi_g$, were found experimentally\cite{Krechkivska2015,Krechkivska2017}.
Their calculated and experimental energy levels indicate a strong vibronic interaction between these two electronic states. 
Figure~\ref{fig:comp_Krechkivska} shows a comparison between the PECs of several $^3\Pi_g$ states calculated in their work and this work. 
The electronic energies of these states are close at short internuclear distances $R<1.4$\,\AA, while the differences increase to about 1200\,cm$^{-1}$ around $R\approx 2.2$\,\AA.
Despite the differences in energy, the shapes of these PECs match well with one another, suggesting that our calculation has good accuracy up to an energy of 40000\,cm$^{-1}$ ($\sim$ 5\,eV) even though a slightly smaller basis set is used in this work.

In summary, based on the results shown here, we have confidence that our calculated PECs have high accuracy in the valence distance range ($R<$1.5\,\AA).
As long as the states are qualititively calculated correctly, the errors in electronic energies should be on the order of 1000\,cm$^{-1}$ and the errors in vibrational frequencies are likely on the order of 20\,cm$^{-1}$.

\subsection{Perturbations and Predissociation of $F\,^1\Pi_u$ state}

The CSE method is applied to study the photodissociation of \ce{C2} in this work.
Diabatic states are more convenient to use as an electronic state basis in the coupled-channel model.
Building the coupled-channel model essentially involves building an interaction matrix $\mathbf{V}(R)$ whose diagonal elements are selected PECs of diabatic electronic states and whose off-diagonal elements are couplings among them, including electrostatic couplings and SOCs.

\begin{figure}
 \centering
 \includegraphics[width=15cm]{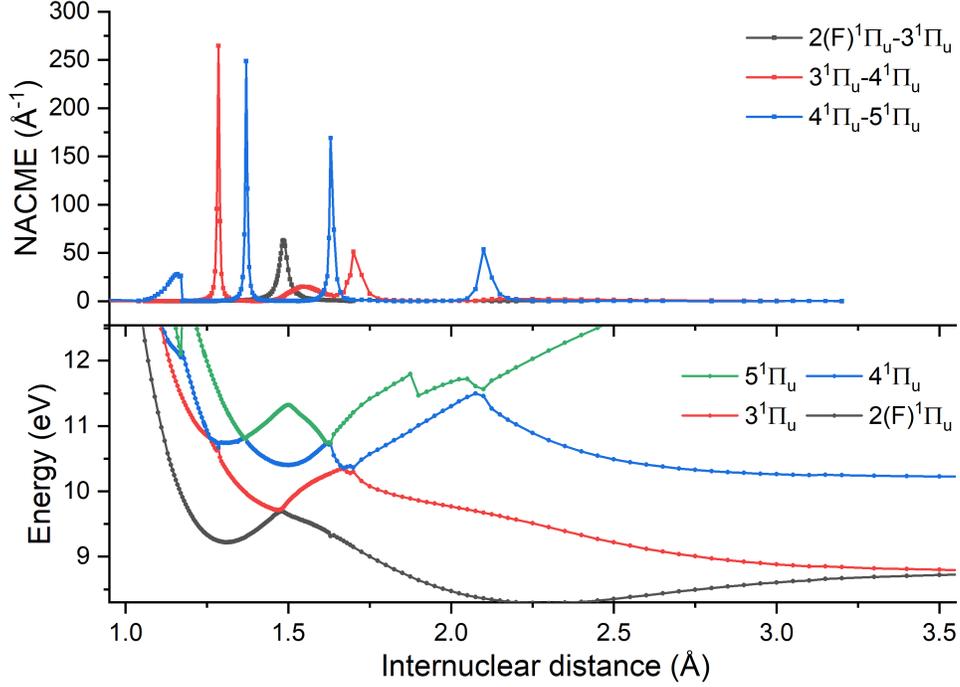}
 \caption{Top: MRCI NACMEs between \ce{C2} $^1\Pi_u$ states. Bottom: MRCI+Q PECs for the same states.}\label{fig:NACME}
\end{figure}

The first step is to build diabatic PECs of $^1\Pi_u$ states from adiabatic ones.
The NACMEs between excited $^1\Pi_u$ states are shown in Figure \ref{fig:NACME}, along with the PECs of these states.
Although it is possible to construct diabatic PECs by applying a unitary  adiabatic-to-diabatic transformation (ADT) which can be calculated mathematically from NACMEs, the frequent nonadiabatic couplings among the $^1\Pi_u$ states makes such a transformation challenging\cite{Nakamura2002}. 
Thus, the NACMEs are only used as a guide to identity where interactions occur.

\begin{figure}
 \centering
 \includegraphics[width=15cm]{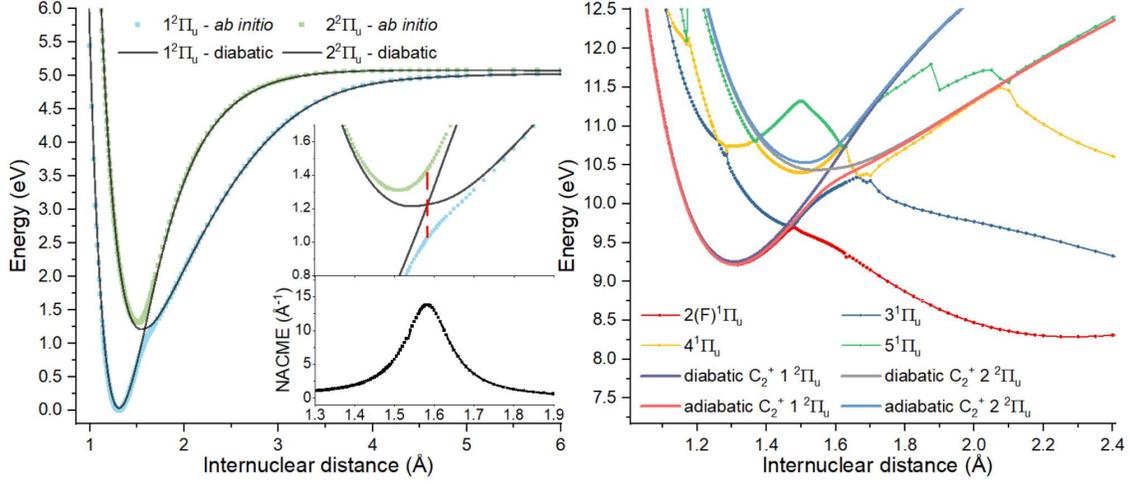}
 \caption{Left: adiabatic and diabatic PECs of low-lying \ce{C2+} $^2\Pi_u$ states. Inset: NACME between the two adiabatic $^2\Pi_u$ states. Right: shifted diabatic \ce{C2+} $^2\Pi_u$ states overlapped with \ce{C2} $^1\Pi_u$ Rydberg states.}\label{fig:ion_diabatic}
\end{figure}

As previously discussed, $F\,^1\Pi_u$ is a Rydberg state with the configuration [$^2\Pi_u,3s$], and so it is expected to have a PEC shape similar to that of the \ce{C2+} $^2\Pi_u$ state.
The MRCI+Q PECs of the two lowest $^2\Pi_u$ electronic states of \ce{C2+} are shown in Figure~\ref{fig:ion_diabatic} (left).
The PEC of the $1\,^2\Pi_u$ state is slightly bent around $R=1.6$\,\AA, indicating it has a nonadiabatic coupling with the $2\,^2\Pi_u$ state.
The potential energy well of the $2\,^2\Pi_u$ state at $R=1.52$\,\AA\ is from this nonadiabatic coupling, instead of an actual potential minimum.
This can be verified by the calculated NACMEs, which show a broad and smooth peak centered at $R=1.6$\AA.
Since this is a simple two-state system, a unitary ADT is used to diabatize these two states. 
We shifted the diabatic PECs of the $1\,^2\Pi_u$ and $2\,^2\Pi_u$ states to make the PEC of adiabatic $1\,^2\Pi_u$ state overlap with the $F\,^1\Pi_u$ state of \ce{C2}, as shown in Figure~\ref{fig:ion_diabatic} (right).
The PEC of the shifted diabatic \ce{C2+} $1\,^2\Pi_u$ state follow the PECs of \ce{C2} $^1\Pi_u$ states closely\cite{Nakamura2002}.
Thus, we use the the shifted PEC of \ce{C2+} $1\,^2\Pi_u$ state to represent the \ce{C2} $F\,^1\Pi_u$ state.
Then, we connect the PECs of $4\,^1\Pi_u$ ($R<1.26$\,\AA), $3\,^1\Pi_u$ ($1.26<R<1.47$\,\AA), and $F\,^1\Pi_u$ ($R>1.47$\,\AA) to build the PEC for a repulsive diabatic $3\,^1\Pi_u$ state.
The electrostatic interaction between these two diabatic states is estimated by half of the energy difference at $R=$1.475\,\AA\ as 0.015\,eV (120\,cm$^{-1}$).
Another two diabatic bound states can be constructed from other $^1\Pi_u$ states.
One corresponds to the \ce{C2+} diabatic $2\,^2\Pi_u$ state.
Its $r_e$ is about 1.6\,\AA, and thus it should not be important for photodissociation studies of \ce{C2} owing to small Franck-Condon factors with the ground $X$ state.
Another is constructed from the $3$, $4$, and $5\,^1\Pi_u$ states with $r_e$ about 1.31\,\AA.
The corresponding TDM is about half of the $F\,^1\Pi_u$ state, and the intensity of absorption from the ground $X$ state is estimated to be one fourth of the $F\,^1\Pi_u - X\,^1\Sigma_g^+$ band.
In this study, we will only focus on the photodissociation via the $F\,^1\Pi_u$ state.
The corresponding TDMs are diabatized by exchanging the curves on both sides of the avoided crossings and interpolating using cubic splines.

\begin{figure}
 \centering
\includegraphics[width=12cm]{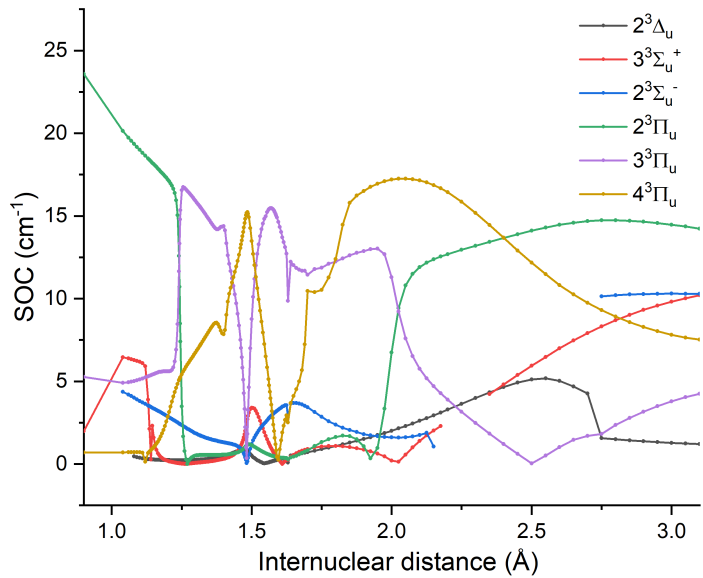}
 \caption{SOCs in cm$^{-1}$ between the $F\,^1\Pi_u$ state of \ce{C2}
 and several nearby triplet states calculated at the MRCI level.}\label{fig:SOC}
\end{figure}

\begin{figure}
 \centering
  \begin{tikzpicture}
   \node[anchor=south west] at (0,0){\includegraphics[width=8cm]{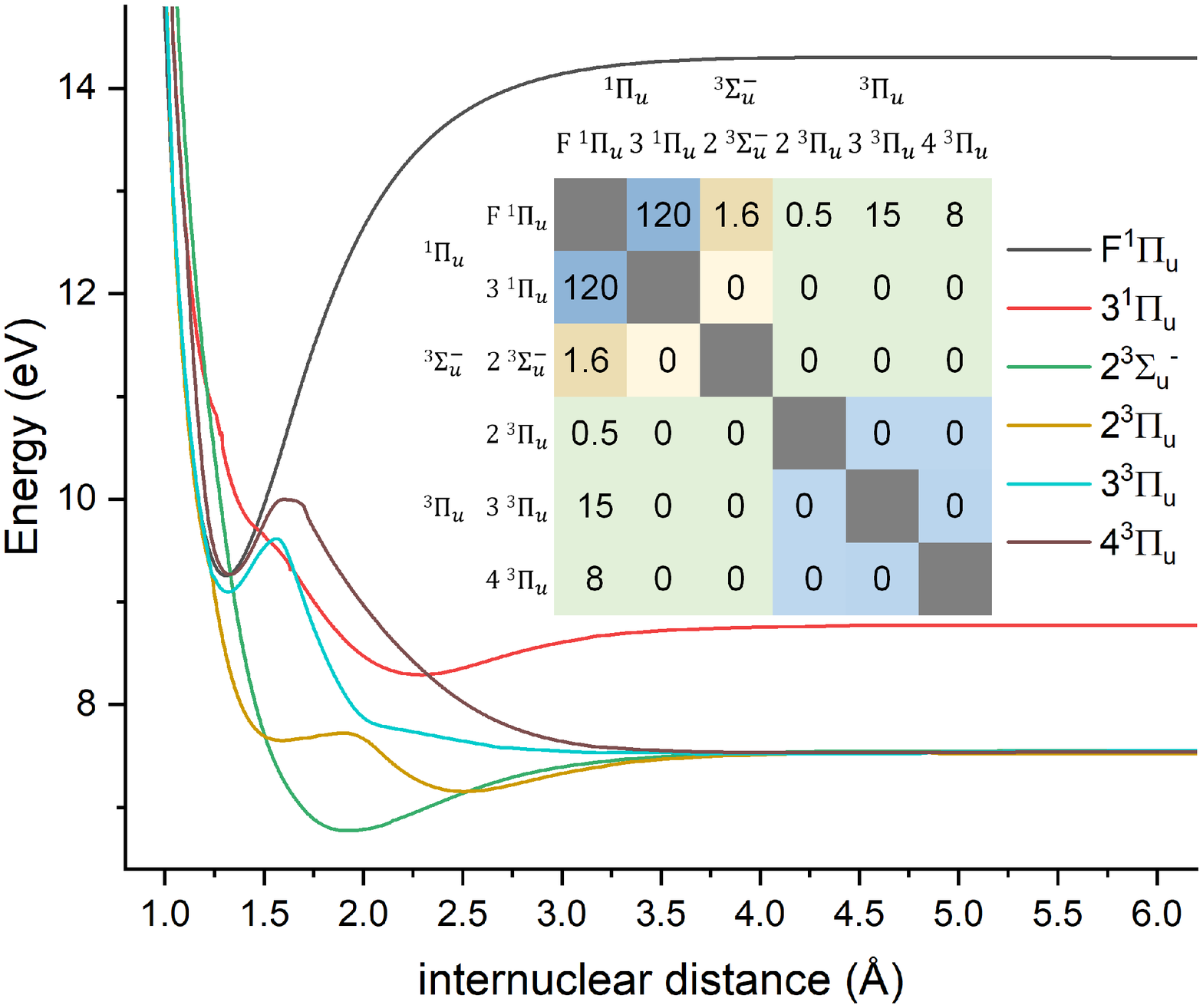}};
   \node[anchor=south west] at (8,0){\includegraphics[width=8cm]{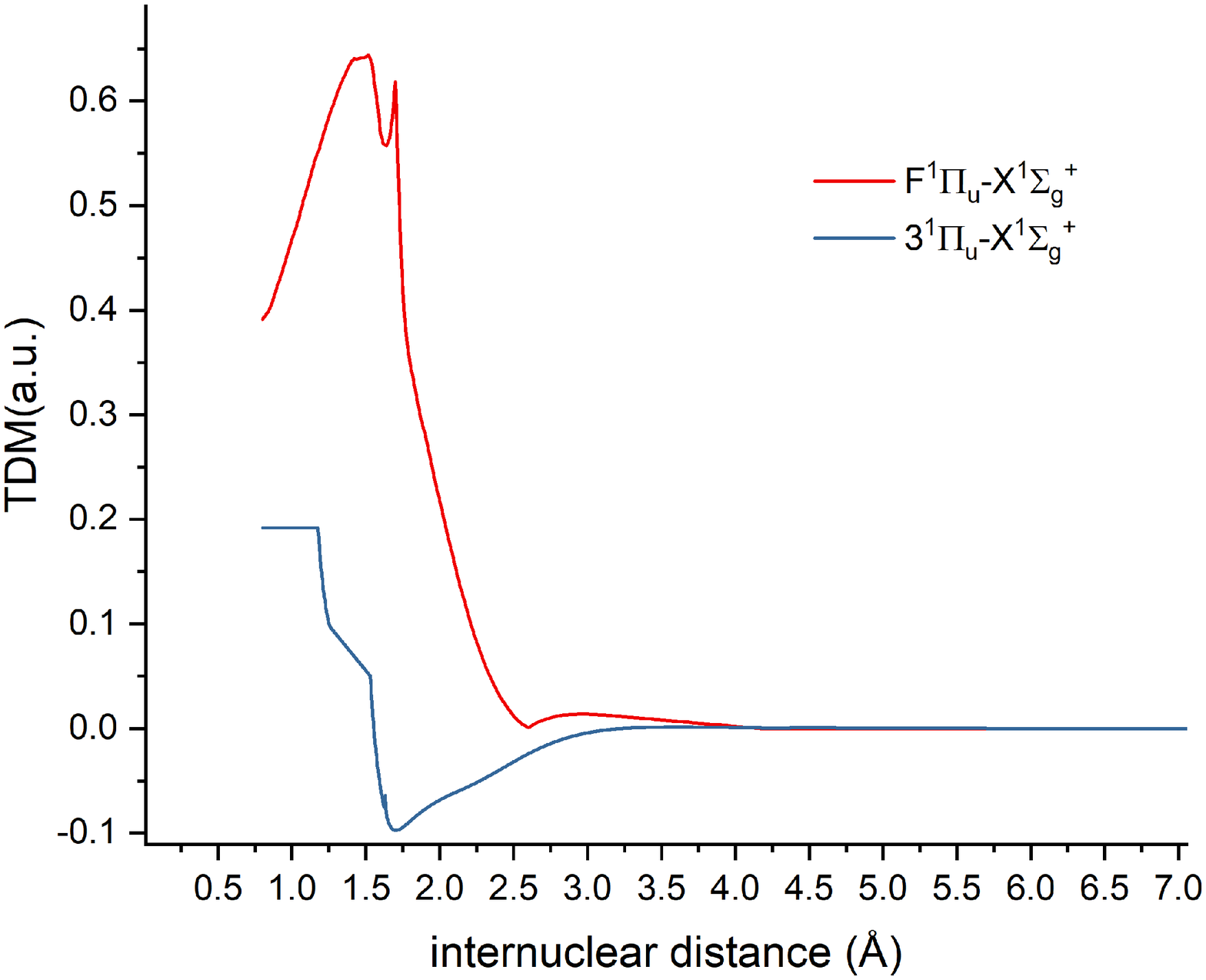}};
  \end{tikzpicture}  
 \caption{Coupled-channel model built for predissociation of the \ce{C2} $F\,^1\Pi_u$ state. Left: PECs of the electronic states, with the interaction matrix inset. Right: TDMs between the ground $X\,^1\Sigma_g^+$ state and diabatic $F\,^1\Pi_u$ and $3\,^1\Pi_u$ states.}\label{fig:couple_model}
\end{figure}

The spin-orbit interaction has been shown to be important in the predissociation of many diatomic molecules, such as \ce{O2}\cite{Lewis2001} and \ce{S2}\cite{Lewis2018}.
Based on the selection rules for spin-orbit coupling which are summarized as
\begin{gather}
	\Delta J = \Delta \Omega = 0;\quad \Delta S = 0,\,\pm 1;\quad \Sigma^+ \leftrightarrow \Sigma^-; \quad g \nleftrightarrow u;  \\
	\Delta \Lambda = \Delta \Sigma =0 \text{ or } \Delta \Lambda = -\Delta \Sigma = \pm 1 \nonumber,
\end{gather}
the $F^2\,\Pi_u$ state is coupled with $^3\Sigma_u^+$, $^3\Sigma_u^-$, $^3\Pi_u$, and $^3\Delta_u$ states.
The PECs of the $3\,^3\Sigma_u^+$, $2\,^3\Sigma_u^-$, and $2\,^3\Delta_u$ states cross that of the $F^1\,\Pi_u$ state at $R=$1.21, 1.33, and 1.21\,\AA, respectively.
In addition, the $2\,^3\Pi_u$, $3\,^3\Pi_u$, and $4\,^3\Pi_u$ states are all close in energy to the PEC of $F^2\,\Pi_u$.
Thus diabatic representations of those states are needed to build the coupled-channel model for predissociation through the $F\,^2\Pi_u$ state.
The calculated SOCs are shown in Figure~\ref{fig:SOC}.
The diabatic PEC of the $3\,^3\Sigma_u^+$ state is constructed similarly to the $3\,^1\Pi_u$ diabatic states: $3\,^3\Sigma_u^+$ converts to $2\,^3\Sigma_u^+$ around $R=$1.53\,\AA, and then to $1\,^3\Sigma_u^+$ around $R=1.75$\,\AA.
Likewise, the diabatic PEC of $2\,^3\Sigma_u^-$ is constructed from the adiabatic PEC of the $2\,^3\Sigma_u^-$ state at $R<$1.67\,\AA\ and the $1\,^3\Sigma_u^-$ state at $R>$1.67\,\AA.
The diabatic PEC of $2\,^3\Delta_u$ crosses with $1\,^3\Delta_u$ around $R=$1.45\,\AA.
However, the SOCs between the $3\,^3\Sigma_u^+$ and $2\,^3\Delta_u$ states with the $F\,^2\Pi_u$ state are almost 0 in the internuclear distance range of 1.2-1.5\,\AA, thus these two states are not considered further.
The coupled-channel model adopted a value of 1.6\,cm$^{-1}$ for the SOC corresponding to $R=$1.33\,\AA\ where a crossing is observed between the $2\,^3\Sigma_u^-$ and $F^1\,\Pi_u$ PECs.
The $3\,^3\Pi_u$ and $4\,^3\Pi_u$ states lie close with each other around $r=$1.25\,\AA, and thus it is challenging to construct diabatic states for them.
In this study, we use the adiabatic curves of the $2\,^3\Pi_u$, $3\,^3\Pi_u$ and $4\,^3\Pi_u$ states as their diabatic representations.
The SOCs between them and the $F^1\,\Pi_u$ state are stable around the equilibrium bond length of the $F^1\,\Pi_u$ state, and so 0.5, 15, 8.0\,cm$^{-1}$ are adopted as the constant SOC value between the $F^1\,\Pi_u$ state and the $2\,^3\Pi_u$, $3\,^3\Pi_u$ and $4\,^3\Pi_u$ states, respectively.
The final coupled-channel model, including the diabatic $F^1\,\Pi_u$, $3^1\,\Pi_u$, $2\,^3\Sigma_u^-$, $2\,^3\Pi_u$, $3\,^3\Pi_u$ and $4\,^3\Pi_u$ states, is shown in Figure~\ref{fig:couple_model}.

\begin{figure}
 \centering
  \includegraphics[width=11cm]{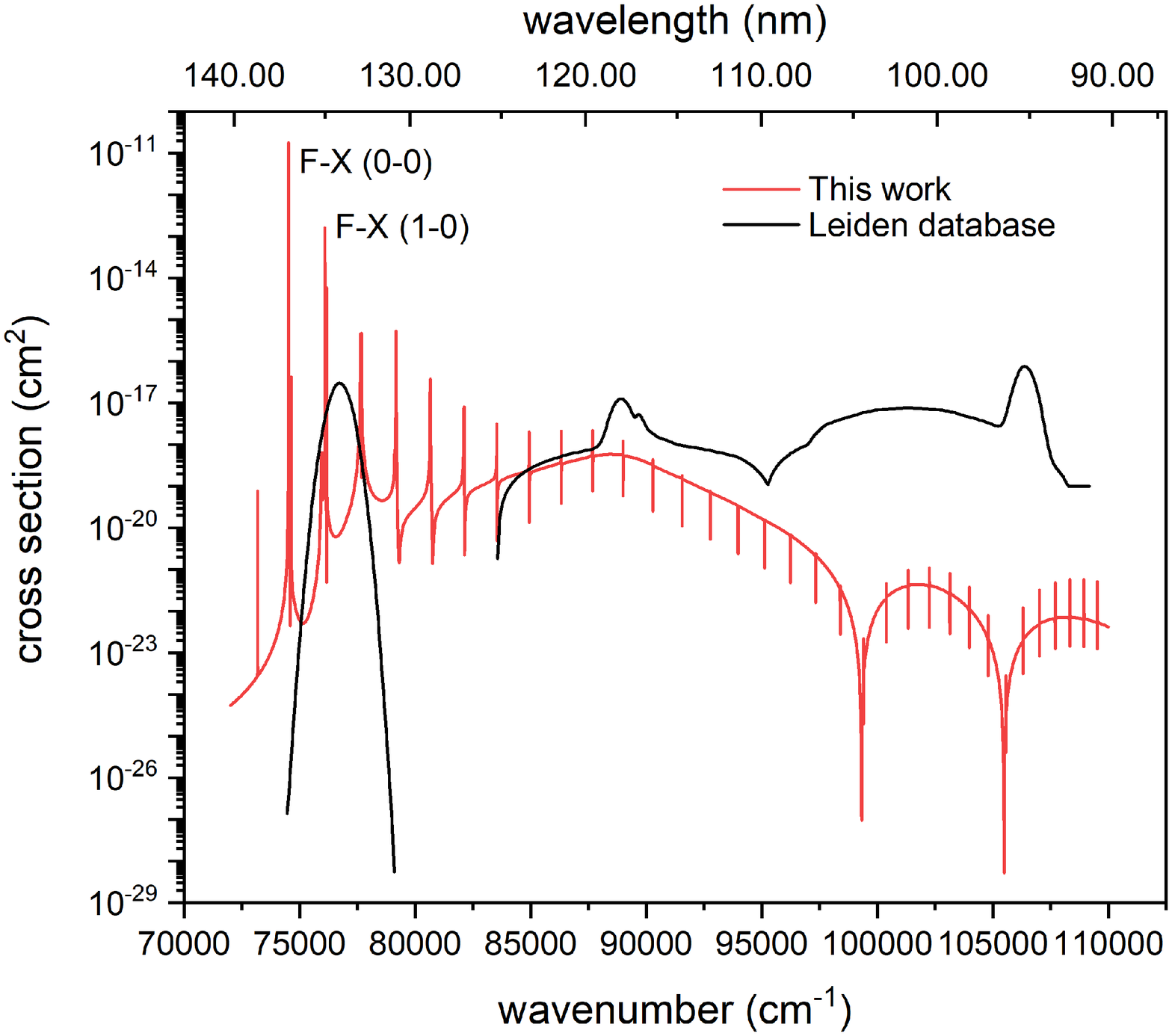}
 \caption{Rotationless photodissociation cross sections of the $X\,^1\Sigma_g^+$ state of \ce{C2} via the $F\,^1\Pi_u$ state.}\label{fig:cs_wavenumber}
\end{figure}

\subsection{Photodissociation cross section and photodissociation rates}

From the coupled-channel model, we calculated the rotationless photodissociation cross section in the energy range between 72000-109600\,cm$^{-1}$ (139-91.2\,nm) from the ground vibronic $X\,^1\Sigma_g^+$ ($v=0$) state via the diabatic $F\,^1\Pi_u - X\,^1\Sigma_g^+$ transition, as shown in Figure~\ref{fig:cs_wavenumber}.
The resolution is 0.1\,cm$^{-1}$ across this range while a smaller interval is used near the $F\,^1\Pi_u - X\,^1\Sigma_g^+$ (0-0) and (1-0) bands.
The current photodissociation cross section adopted in the Leiden database~\cite{Heays2017} is presented for comparison.
In the Leiden database photodissociation cross section curve, the peak around 134\,nm (74600\,cm$^{-1}$) is based on the previous experimental $F-X$ bands~\cite{Herzberg1969a}, and the double peaks around 118\,nm (84700\,cm$^{-1}$) are derived from previous theoretical results of $F-X$ bands~\cite{Pouilly1983}.
The linewidths are assumed to be 1\,nm.
Our calculated $F-X$ (0-0) band is located at 74521.2\,cm$^{-1}$ with a linewidth of 0.0014\,cm$^{-1}$ and an integrated cross section of 7.98$\times$10$^{-14}$\,cm$^2$cm$^{-1}$, while the $F-X$ (1-0) band is at 76099.7\,cm$^{-1}$ with a linewidth of 0.079\,cm$^{-1}$ and integrated cross section of 4.00$\times$10$^{-14}$\,cm$^2$cm$^{-1}$.
The derived predissociation lifetime $\tau_{pd}$ is 3.78\,ns and 0.067\,ns for the $F\,^1\Pi_u$ $v=0$ and $v=1$ levels.
The first potential well of the adiabatic $F\,^1\Pi_u$ state is about 3800\,cm$^{-1}$ deep and is barely able to support the diabatic $v=2$ vibrational level.
Nevertheless, vibrational levels above $v^{\prime}=1$ are unlikely to contribute significantly to photodissociation because they have small Franck–Condon factors.
The photodissociation cross section curve beyond the $F-X$ (1-0) band is not thought to be accurate from our model, because no contributions from electronic states above the $F$ state are included.
Nevertheless, the peak at 118\,nm in the Leiden curve is likely unphysical as discussed above.

Spontaneous emission lifetimes ($\tau_{rad}$) of the $F\,^1\Pi_u$ $v=0$ and $v=1$ levels are computed from the adiabatic PECs and TDMs shown in Figure~\ref{fig:PECs} and Figure~\ref{fig:TDMs_2} using the program DUO\cite{Yurchenko2016}.
As discussed above, only the $F\,^1\Pi_u - X\,^1\Sigma_g^+$ and $F\,^1\Pi_u - B\,^1\Delta_g$ transitions are considered.
The total Einstein $A_{21}$ coefficient is computed from
\begin{equation}
  A_{21} (F, v^{\prime}) = \sum_{v^{\prime\prime}} A_{21} (F-X,v^{\prime}-v^{\prime\prime}) + \sum_{v^{\prime\prime}} A_{21} (F-B,v^{\prime}-v^{\prime\prime}).
\end{equation}
The $A_{21}$ coefficients for the $F\,^1\Pi_u - X\,^1\Sigma_g^+$ transition are 2.95$\times$10$^{8}\,$s$^{-1}$ for the $v=0$ level and 2.92$\times$10$^{8}\,$s$^{-1}$ for the $v=1$ level, while for the $F\,^1\Pi_u - B\,^1\Delta_g$ transition they are 5.34$\times$10$^{7}\,$s$^{-1}$ for the $v=0$ level and 5.28$\times$10$^{7}\,$s$^{-1}$ for the $v=1$ level.
The values of the $A_{21}$ coefficients yield a total lifetime of 2.87\,ns for the $v=0$ level and 2.90\,ns for the $v=1$ level.
Based on our model, the predissociation through the $F\,^1\Pi_u$ $v=1$ level is more than 40 times faster than spontaneous emission, while the predissociation via its $v=0$ level is a little slower than spontaneous emission, as summarized in Table~\ref{tab:C2_lines}.
Based on the calculated predissociation and spontaneous emission lifetimes, 43.1\% of photoabsorption would give rise to predissociation.

As a comparison, a lifetime of 0.006\,ns for both the $F\,^1\Pi_u$ $v=0$ and $v=1$ levels was derived from measured linewidths of the $F\,^1\Pi_u - X\,^1\Sigma_g^+$ (0-0) and (1-0) transitions, which suggests the $F\,^1\Pi_u$ state decays rapidly via predissociation\cite{Hupe2012}.
This number is significantly smaller than our predissociation lifetimes of 3.78 and 0.067\,ns for the $F\,^1\Pi_u$ $v=0$ and $v=1$ levels, respectively.
The disagreement may come from either an overestimation of the resolution of the astronomical observations or inaccuracies in our coupled-channel model.
However, a combined analysis of the resolution of the observations \cite{Sheffer2007,Hupe2012} suggests that the inferred lifetime of 0.006\,ns should be reliable to $\pm$25\%, suggesting the cause of disagreement lies with the calculations. 
The accuracy of our calculated linewidths and lifetimes depends sensitively on the calculated \textit{ab initio} couplings.
For example, if the SOC between the $F\,^1\Pi_u$ and $2\,^3\Sigma_u^-$ states increases from 1.6\,cm$^{-1}$ to 10\,cm$^{-1}$, then the calculated linewidth for $F-X$ (0-0) increases from 0.0014\,cm$^{-1}$ to 0.050\,cm$^{-1}$, which corresponds to a lifetime of 0.11\,ns.
The predissociation in our model would then be significantly faster than spontaneous emission.
However,  even larger corrections need to be applied to match the measured lifetime of 0.006\,ns.

\begin{table}[hbtp]
\centering
\begin{threeparttable}
\setlength\extrarowheight{2pt}
\caption{Properties of the $F-X$ transitions of \ce{C2}.\label{tab:C2_lines}}
\begin{tabularx}{\textwidth}{@{\extracolsep{\fill}}*{11}{c}}
    \toprule
    \multicolumn{2}{c}{Band} & $v_{\text{expt}}$  \tnote{a} & v  & $\gamma$  & $\sigma_0$  & $\tau_{pd}$ & $A_{21,(F-X)}$ & $A_{21,(F-B)} $ & $A_{21,tot}$  & $\tau_{rad}$  \\
    & & (cm$^{-1}$) & (cm$^{-1}$) & (cm$^{-1}$) & (cm$^2$cm$^{-1}$) & ns & (s$^{-1}$) & (s$^{-1}$) & (s$^{-1}$) &(ns) \\
    \midrule
    $F-X$ & (0-0) & 74550\tnote{c}  & 74521.2 & 0.0014 & 7.98\ee{-14} & 3.78 & 2.95\ee{8} & 5.34\ee{7} & 3.48\ee{8} & 2.87   \\
          & (1-0) & 76105\tnote{c}  & 76099.7 & 0.079 & 4.00\ee{-14} & 0.067 & 2.92\ee{8} & 5.28\ee{7} & 3.45\ee{8} & 2.90  \\
    \bottomrule
\end{tabularx}
\begin{tablenotes}
    \item{a.} Band heads~\cite{Herzberg1969a}.
\end{tablenotes}
\end{threeparttable}
\end{table}

In our current model, the branching ratios can be obtained by comparing the photodissociation cross sections of different open channels.
Predissociation of both $v=0$ and $v=1$ levels produces $^3P+{^1D}$ atomic carbon products since all possible predissociation pathways through triplet states converge to this atomic limit.

Assuming all photoabsorption leads to photodissociation, under the standard interstellar radiation field (ISRF), the photodissociation rate over the range of wavenumbers 72000-80000\,cm$^{-1}$ is 5.02$\times$10$^{-10}$\,s${^{-1}}$, with a contribution from the $F-X$ (0-0) transition of 2.94$\times$10$^{-10}$\,s${^{-1}}$ and from the $F-X$ (1-0) transition of 1.39$\times$10$^{-10}$\,s${^{-1}}$.
If the 43.1\% photodissociation efficiency is applied for the $v=0$ level, then the corresponding rate from the $F-X$ (0-0) band would be 1.27$\times$10$^{-10}$\,s${^{-1}}$, giving a total photodissociation rate of 3.35$\times$10$^{-10}$\,s${^{-1}}$.
The photodissociation rate arising from wavenumbers above 80000\,cm$^{-1}$ is only 1.83$\times$10$^{-11}$\,s${^{-1}}$, which is negligible compared to the (0-0) and (1-1) transitions.
As explained above, the Leiden database contains two transitions involving the $F$ state.
One is in the wavelength range 130-134\,nm, which is likely from the Herzberg $F-X$ band, and the other, in the range 115-120\,nm, is likely from the $F-X$ transition in a previous theoretical study\cite{Pouilly1983}.
The photodissociation rates calculated from these two bands are 6.66$\times$10$^{-11}$\,s${^{-1}}$ and 3.59$\times$10$^{-11}$\,s${^{-1}}$, respectively, and the total photodissociation rate for \ce{C2} in ISRF is 2.35$\times$10$^{-10}$\,s${^{-1}}$.
The calculated photodissociation rate from the $F-X$ bands even assuming a reduced photodissociation efficiency is still larger than the total photodissociation rate in the Leiden database. 
This is likely due to the low oscillator strength of the $F-X$ bands ($f_{00}=0.02$) derived from the theoretical calculation on which the Leiden database cross section are based.
Notably, both previous astronomical observations \cite{Lambert1995,Sonnentrucker2007} and theoretical calculations \cite{Bruna2001} have also derived much larger oscillator strengths comparable to those calculated here.
Thus it is likely that the photodissociation rate under the ISRF is underestimated by present astronomical models.




\section{Conclusion\label{sec:C2_theo_conclusion}}

Here we have presented a detailed \textit{ab initio} theoretical study of \ce{C2} photodissociation focusing on the predissociation of the $F\,^1\Pi_u$ state.
Potential energy curves for a total of 57 electronic states were calculated with the DW-SA-CASSCF/MRCI+Q method with a basis set aug-cc-pV5Z+2s2p.
By using an (8,12) active space, the Rydberg nature of the $F\,^1\Pi_u$ level was confirmed, and non-adiabatic couplings among the excited $^1\Pi_u$ states as well as SOCs between $F\,^1\Pi_u$ and other triplet states were explored.
Then, a coupled-channel model was used to simulate the photodissociation cross section of \ce{C2} via its $F\,^1\Pi_u - X\,^1\Sigma_g^+$ transition.

We reproduced the $F\,^1\Pi_u - X\,^1\Sigma_g^+$ (0-0) and (1-0) bands in our photodissociation cross section calculation.
By comparing the derived predissociation lifetime with the computed spontaneous emission lifetime, the $v=1$ level was found to decay rapidly through predissociation.
Unlike the results reported by\cite{Hupe2012}, the predissociation rate of the $v=0$ level was found to be comparable with spontaneous emission in this study.
Accurate modeling of predissociation depends on precise coupling terms, which would benefit from further experimental studies of the $F-X$ band.
Moreover, we predict a strong $2\,^1\Sigma_u^+ - X\,^1\Sigma_g^+$ absorption peak around 10.7\,eV (115.9\,nm) which could also give rise to fast predissociation, and should be a priority for experimental measurements.

\section{Supplementary Material\label{sec:SI}}

See supplementary material for PECs, TDMs and photodissociation cross sections.
The data are also available in machine-readable format.

\begin{acknowledgments}
This work was supported by the NASA Astrophysics Research and Analysis program under awards 80NSSC18K0241 and 80NSSC19K0303.
\end{acknowledgments}

\end{document}